\documentclass{aa}
\usepackage{longtable}

\usepackage{psfig}
\usepackage{supertabular}

\begin{document}
% ====================================================================
%                       MACROS PERSO
% =====================================================================
%
% -----
% Abrev
% -----
\def\lfir{$L_{\rm FIR}$}
\def\mabs{M$_{\rm abs}$}
\def\etal{et al.}
\def\hii{H{\sc ii}}
\def\cbeta{$c_{\rm H\beta}$}
\def\av{A$_{\rm v}$}
\def\flam{$F_{\lambda}$}
\def\ilam{$I_{\lambda}$}
% -----
% Units
% -----
\def\micron{$\mu$m}
\def\kms{km s$^{-1}$}
\def\kmsmpc{km s$^{-1}$ Mpc$^{-1}$}
\def\cmc{cm$^{-3}$}
\def\erg{ergs s$^{-1}$ cm$^{-2}$ \AA$^{-1}$}
\def\ergs{ergs s$^{-1}$}
\def\ergscm{ergs s$^{-1}$ cm$^{-2}$}
\def\lsun{L$_{\odot}$}
\def\msun{M$_{\odot}$}
\def\zsun{Z$_{\odot}$}
% -----
% Lines
% -----
% Hydrogen
%
\def\halpha{\ifmmode {\rm H{\alpha}} \else $\rm H{\alpha}$\fi}
\def\hbeta{\ifmmode {\rm H{\beta}} \else $\rm H{\beta}$\fi}
%
% Helium
%
\def\heia{He\,{\sc i} $\lambda$4471}
\def\heib{He\,{\sc i} $\lambda$4922}
\def\heic{He\,{\sc i} $\lambda$5876}
\def\heiia{He\,{\sc ii} $\lambda$4686}
\def\heiib{He\,{\sc ii} $\lambda$5412}
\def\heii{\ifmmode {{\rm He{\sc ii}} \lambda 4686} \else {He\,{\sc ii} $\lambda$4686}\fi}
\def\Heii{He\,{\sc ii}}

% Oxygen
%
\def\oia{[O\,{\sc i}] $\lambda$6300}
\def\oib{[O\,{\sc i}] $\lambda$6364}
\def\oii{[O\,{\sc ii}] $\lambda$3727}
\def\oiii{[O\,{\sc iii}]}
\def\oiiia{[O\,{\sc iii}] $\lambda$4959}
\def\oiiib{[O\,{\sc iii}] $\lambda$5007}
\def\ov{O\,{\sc v} $\lambda$5590}
%
% Nitrogen
%
\def\ni{[N\,{\sc i}] $\lambda$5199}
\def\nii{[N\,{\sc ii}] $\lambda$5755}
\def\niii{N\,{\sc iii} $\lambda$4640}
\def\nv{N\,{\sc v} $\lambda$4604}
%
% Carbon
%
\def\ciii{C\,{\sc iii} $\lambda$4650}
\def\ciiia{C\,{\sc iii} $\lambda$4658}
\def\ciiib{C\,{\sc iii} $\lambda$5696}
\def\civ{C\,{\sc iv} $\lambda$5808}
%
% Sulfur
%
\def\sii{[S\,{\sc ii}]}
\def\siia{[S\,{\sc ii}] $\lambda$6716}
\def\siib{[S\,{\sc ii}] $\lambda$6731}
\def\Sii{[S\,{\sc ii}] $\lambda\lambda$6716,6731}
\def\siii{[S\,{\sc iii}] $\lambda$6318}
%
% Argon
%
\def\ariva{[Ar\,{\sc iv}] $\lambda$4711}
\def\arivb{[Ar\,{\sc iv}] $\lambda$4740}
%
% Iron
%
\def\feiiia{[Fe\,{\sc iii}] $\lambda$4658}
\def\feiiib{[Fe\,{\sc iii}] $\lambda$5271}
%
% Chlore
%
\def\cliiia{[Cl\,{\sc iii}] $\lambda$5518}
\def\cliiib{[Cl\,{\sc iii}] $\lambda$5538}
%
% Line ratios
%
\def\oiiishb{[O\,{\sc iii}]/H$\beta$}
\def\niisha{[N\,{\sc ii}]/H$\alpha$}
\def\siisha{[S\,{\sc ii}]/H$\alpha$}

% references
%
\def\aap{A\&A}
\def\aas{A\&AS}
\def\aj{AJ}
\def\apj{ApJ}
\def\apjl{ApJ}
\def\apjs{ApJS}
\def\mnras{MNRAS}
\def\pasp{PASP}

% NOTE: the \def command does not allow for numbers in the names !
%       It will just recognize the name up to the first number...

\def\asp35{in ``Massive Stars: Their Lives in the Interstellar Medium'',
        Eds. J.P. Cassinelli, E.B. Churchwell, ASP Conf. Series, 35}

\def\iau105{in ``Observational Tests of the Stellar Evolution Theory'', 
        Eds. A. Maeder, A. Renzini, IAU Symp. 105}

\def\iaugreece{in ``Luminous Stars and Associations in Galaxies'', Eds. C. Loore,
        A.J. Willis, P. Laskarides, IAU Symp. 116, Kluwer:  Netherland}

\def\iaubali{in ``Wolf-Rayet Stars and Interrelations with Other Massive
        Stars in Galaxies'', Eds. K.A. van der Hucht, B. Hidayat, IAU Symp. 142, 
        Kluwer: Netherland}
\def\liege{in ``WR Stars in the Framework f Stellar Evolution'', 33rd Li\`ege Int. 
        Astroph. Coll., Eds. J.M. de Vreux et al. (Li\`ege: Universit\'e de Li\`ege)}
\def\iaumex{in ``Wolf-Rayet Phenomena in Massive Stars and Starburst Galaxies'', 
        K.A. van der Hucht, G. Koenigsberger, \& P.R.J. Eenens (eds.), 
        IAU Symp. 193, (San Francisco: ASP)}
\def\crete{in ``From Stars to Galaxies: The Impact of Stellar Physics on Galaxy Evolution'',
  Eds. C. Leitherer, U. Fritze-v.\ Alvensleben, J. Huchra, ASP Conf. Series, 98}

\def\yt98{Izotov \& Thuan (1998)}
\def\y94{Izotov et al.\ (1994)}
\def\yu97{Izotov et al.\ (1997a)}
\def\git{Guseva et al.\ (1998)}

\def\kj85{Kunth \& Joubert (1985)}
\def\vc92{Vacca \& Conti (1992)}

%%%%%%%%%%%%%%%%%%%%%%%%%%%%%%%%%%%%%%%%%%%%%%%%%%%%%%%%%%%%%%%%%%%%%%%%
%\begin{document}

\thesaurus{} 

\title{New catalogue of Wolf-Rayet galaxies and high-excitation 
extra-galactic \hii\ regions
%\thanks{}
}

\author{Daniel Schaerer \inst{1} \and
           Thierry Contini  \inst{2,3} \and Maximilien Pindao \inst{4}}

\offprints{D. Schaerer, schaerer@obs-mip.fr}

\institute{
Observatoire Midi-Pyr\'en\'ees, Laboratoire d'Astrophysique, 14, Av. E. Belin, 
F-31400 Toulouse, France
\and
School of Physics \& Astronomy, Tel Aviv University, 69978 Tel Aviv, Israel
\and
European Southern Observatory, Karl-Schwarzschild-Str.\ 2, D-85748 Garching 
bei M\"unchen, Germany
\and
Observatoire de Gen\`eve, 51, Ch.\ des Maillettes, CH-1290 Sauverny, Switzerland
}

\date{Received 27 October 1998 / Accepted 24 November 1998}
%\date{\bf Submitted to A\&A}

\titlerunning{Wolf-Rayet galaxies and high-excitation \hii\ regions}
\authorrunning{Schaerer et al.}
\maketitle

%%%%%%%%%%%%%%%%%%%%%%%%%%%%%%%%%%%%%%%%%%%%%%%%%%%%%%%%%%%%%%%%%%%%%%%%
\begin{abstract} 
We present a new compilation of Wolf-Rayet (WR) galaxies
and extra-galactic \hii\ regions showing {\em broad} \heii\ emission
drawn from the literature.
Relevant information on the presence of other broad emission lines (\niii, \civ\ and 
others) from WR stars of WN and WC subtypes, and other existing broad 
nebular lines is provided.

%dsnumber TOTAL, CIV
In total we include 139 known WR galaxies. Among these, 57 objects show 
both broad \heii\ and \civ\ features. In addition to the broad (stellar) 
\heii\ emission, a {\em nebular} \Heii\ component is well established
%dsnumber next 2 lines
(suspected) in 44 (54) objects.
We find 19 extra-galatic \hii\ regions without WR detections showing 
nebular \heii\ emission.
%, although this sample may be less complete.

The present sample can be used for a variety of studies on massive stars, 
interactions of massive stars with the ISM, stellar populations, starburst
galaxies etc.\
The data is accessible electronically and will be updated periodicaly.


\end{abstract}

% ---------------------------------------------------------------------------
%

%%%%%%%%%%%%%%%%%%%%%%%%%%%%%%%%%%%%%%%%%%%%%%%%%%%%%%%%%%%%%%%%%%%%%%%%
\section{Introduction}
Wolf-Rayet (WR) galaxies are extragalactic objects whose integrated
spectra show direct signatures from WR stars, most commonly a broad
\heii\ feature originating in the stellar winds of these stars.
Since the first detection of such a feature in He 2-10 (Allen \etal\ 1976),
a large number of WR galaxies have been reported, some in systematic
searches (e.g.\ Kunth \& Joubert 1985), but mostly serendipitously.
For example many objects with WR features have been found in samples
of high S/N spectra of low metallicity extra-galactic \hii\ regions 
aimed at deriving the primordial He abundance (cf.\ Izotov \& Thuan 1998).
It must be reminded to use the term WR ``galaxy'' with caution.
Depending e.g.\ on the distance of the object and the spatial extension 
of the observation, the region of concern may be ``just'' a single
extra-galactic \hii\ region with a few WR stars in a galaxy or the 
nucleus of a powerful starburst galaxy harbouring numerous massive
stars.

Since the compilation of Conti (1991) listing 37 objects, the number 
of known WR galaxies has grown rapidly to more than 130 in the present
%dsnumber TOTAL
catalogue.
Interestingly many objects are now found showing additional features
from WR stars in their spectra. E.g.\ the broad emission lines of \niii\ 
and/or \ciii\ as well as \civ, among the strongest optical lines in WN and 
WC stars, are increasingly often being detected.
Lines originating from WC stars (representing more evolved phases than WN stars)
provide useful independent and complementary information on the massive star
content in these regions (e.g.\ Schaerer \etal\ 1999).
 
By definition it is not surprising that WR galaxies do not form
a homogeneous class. Indeed, WR galaxies are found among a large variety
of morphological types, from low-mass blue compact dwarf (BCDs) and 
irregular galaxies, to massive spirals and luminous merging IRAS galaxies. 
Recent studies also quite convincingly show the evidence
of signatures from WR stars in Seyfert 2 and LINERs (Osterbrock \& Cohen 1982, 
Ho et al\ 1995, Heckman \etal\ 1997, Storchi-Bergmann \etal\ 1998, 
Kunth \& Contini 1998).
Allen (1995) claims even the possible detection of WR stars in 
central cluster galaxies of two cooling flows out to a redshift of $z \sim$ 0.25.

Empirically all WR galaxies show nebular emission lines.
The absolute scales (absolute magnitudes, ionizing fluxes etc.) of the investigated
objects vary greatly; generally speaking the properties of WR galaxies overlap 
with those of other emission line galaxies and form a continuous extension of 
giant \hii\ regions (Conti 1991).

For most ``traditional'' WR galaxies (e.g.\ \hii\ galaxies, BCDs etc.) 
the nebular spectrum is likely due to photoionization of stellar origin.
However, this statement does obviously not hold in general, e.g.\ for Seyfert 2 
and LINER revealing the presence of WR stars.
Among the former ``class'' a considerable fraction ($\sim$ 1/3 in the present 
compilation) of objects also show {\em nebular} \heii\ emission in addition
to the {\em broad} WR feature.  
This line is also present in some giant \hii\ regions where no WR features 
have been detected.
Except in planetary nebulae, nebular \heii\ emission is very rarely found in 
Galactic \hii\ regions (cf.\ Garnett \etal\ 1991, Schaerer 1997).
Its origin, requiring sources with sufficient photons of energy $>$ 54 eV, 
has remained puzzling (see Garnett \etal\ 1991 and references therein).
Supported by quantitative modeling, Schaerer (1996) has suggested that the
origin of nebular \heii\ emission is intimately linked with the appearance of hot
WR stars.
To facilitate systematic analysis on the origin of {\em nebular} \heii\ emission
we therefore also include the relevant information on objects showing this line.
Such studies have a bearing on our understanding of physical processes in \hii\ 
regions and related nebulae, the ionizing fluxes of starbursts and their contribution
to the ionization of the intergalactic medium etc. (cf.\ Garnett \etal\ 1991,
Schaerer \etal\ 1998, Stasi\'nska 1998).
%Stasi\'nska \& Schaerer 1998).

The minimum common property of all WR galaxies is (provided our the origin of 
the considered line and our understanding of stellar evolution is correct) 
{\em ongoing or recent star formation} which has produced stars
massive enough to evolve to the WR stage.
This indicates typically ages of $\la$ 10 Myr and stars with initial
masses $M_{\rm ini} \ga$ 20 \msun\ (Maeder \& Conti 1994).

WR galaxies are therefore ideal objects to study the early phases of 
starbursts, determine burst properties (age, duration, SFR), and to constrain
parameters (i.e.\ slope and upper mass cut-off) of the upper part of the 
initial mass function (see e.g.\ Arnault \etal\ 1989, Mas-Hesse \& Kunth 1991, 
1998, Kr\"uger \etal\ 1992, Vacca \& Conti 1992, Meynet 1995, Contini
\etal\ 1995, Schaerer 1996, Schaerer \etal\ 1999).
Conversely studies of the stellar populations in super star clusters
frequently formed in starbursts and WR galaxies (Conti \& Vacca 1994, Meurer
\etal 1995) can also place constraints on stellar evolution models for
massive stars, e.g.\ at extremely low metallicities, which are inaccessible in the
Local Group (cf.\ I Zw 18: Izotov \etal\ 1997b, Legrand \etal\ 1997, de Mello 
\etal\ 1998).

As galaxies exhibiting intense star formation are being discovered in 
large numbers at progressively larger distances, ``template''
systems become increasingly important for our understanding of
distant objects which cannot be studied to the same depth.
As such WR galaxies represent useful templates of young 
starbursts which show close resemblance to recently discovered high redshift 
galaxies (cf.\ Leitherer \etal\ 1996, Ebbels \etal\ 1996, Lowenthal \etal\ 1997).

The present compilation should facilitate future systematic studies on some
of the issues discussed above.
The structure of the paper is as follows.
In Sect.\ \ref{s_search} we review the searches undertaken up to date for 
WR galaxies.
In Sect.\ \ref{s_catalogue} the compilation of all known WR galaxies is
presented. Brief remarks on each individual object are given in Sect.\ \ref{s_objects}.
The list of extra-galactic \hii\ regions showing only {\em nebular} \heii\
is given in Sect.\ \ref{s_heii}.
Suspected WR galaxies are discussed in Sect.\ \ref{s_cand}.
A brief discussion and our main conclusions are found in Sect.\ \ref{s_conclude}.

%%%%%%%%%%%%%%%%%%%%%%%%%%%%%%%%%%%%%%%%%%%%%%%%%%%%%%%%%%%%%%%%%%%%%%%%
\section{Searches for WR galaxies}
\label{s_search}
Few systematic searches for WR populations outside 
the Local Group (or ``WR galaxies'') have been undertaken.
In this Section we briefly summarise the studies explicitly devoted
to the detection of WR signatures. 
A list of candidate WR galaxies resulting from some of these searches 
or found loosely in the literature is provided in Sect.\ \ref{s_cand}.
The WR galaxies issued from the searches described below are included
in our list and represent the vast majority of detections. 
Let us now briefly summarise the properties of the spectroscopic 
and narrow-band imaging searches.

\subsection{Spectroscopic searches}
The first search for WR features in giant \hii\ regions of nearby galaxies
was carried out by D'Odorico \etal\ (1983). The latest update from their
study is summarised by Rosa \& D'Odorico (1986).
Data from their work was included in the quantitative analysis by
Arnault \etal\ (1989).

The most detailed search was undertaken by Kunth \& Joubert (1985)
from a sample of 45 ``lazy'' galaxies (blue emission-line galaxies forming
stars by intermittent short bursts) from various sources. 
In their statistical approach they measure the excess emission above the
continuum between 4600-4711 \AA\ (rest wavelength) after subtracting
a typical nebular contamination taken as a function of the excitation
level and abundance. Their search yielded 19 regions (15 different objects)
with excess emission above 0.8 $\sigma$.

A systematic search for a broad WR bump in all the \hii\ galaxies 
included in the catalogue of Terlevich \etal\ (1991) was presented by 
Masegosa \etal\ (1991).
Earlier publications using a subset of the same observational data had
also reported some WR detections and nebular \heii\ (Campbell \& Smith 1986,
Campbell \etal\ 1986).
Positive detections were considered by Masegosa \etal\ when the ``blue bump'' 
was at least 1 $\sigma$ over the continuum level and clearly discernible 
from the nebular \heii\ line. Their search yielded 37 detections ($\sim $ 10\%
of the sample); 14 of these objects have spectra with a spectral resolution of 
$\la$ 5 \AA\ FWHM, which the authors estimate to be ``good enough''
to reliably detect WR stars.
Only these objects (their Table 2) were included in our list as confirmed 
WR galaxies. The remaining objects are classified here as ``candidates''
(Sect.\ \ref{s_cand}).

Recently Pindao (1998) and Pindao \etal\ (1999) have reanalysed the spectra 
from the Terlevich \etal\ catalogue and $\ga$ 100 additional emission line 
galaxies for their WR content.
Objects with a clear detection of broad \heii\ are retained as WR galaxies here
(See Pindao 1998. The detection level corresponds to $\ga$ 0.8 $\sigma$, i.e.\
category 4 of Pindao \etal\ 1999).
Category 3 objects (WR bump detection at $\sim$ 0.5 $\sigma$) 
from Pindao \etal\ (1999) are classified here as ``WR candidates''.
%Objects from their category 4 (clear detection of broad \heii\ above $\sim$
%0.8 $\sigma$) are retained as WR galaxies here; category 3 objects (WR bump
%detection between $\sim$ 0.5--0.8 $\sigma$) are classified here as ``WR candidates''.

Robledo-Rella \& Conti (1993) presented a search for WR features
in a selected sample of northern \hii\ galaxies; candidates are
given in Sect.\ \ref{s_cand}.

First results from a new search for WR signatures in young starbursts
have been presented by Contini (1996) and Kovo \& Contini (1998).

An ongoing systematic search for WR galaxies has been mentioned by Huang 
\etal\ (1998).

According to Izotov (1998, private communication) the observational data
gathered primarily for accurate determinations of the helium abundance
since 1993 (see Izotov \etal, 1994, 1996, 1997a; Thuan \etal\ 1995, Izotov \& 
Thuan 1998) are being systematically re-analysed for their WR content 
(Guseva \etal\ 1998, Izotov \etal\ 1998). 
Adding 10 newly observed objects, their sample mostly including blue compact 
galaxies consists of $\sim$ 70 spectra. 
While the initial sample contained essentially very metal-poor 
objects, metallicities up to $\sim$ solar are now also included.
The majority of the WR detections have been mentioned in the above papers;
in total \git\ and Izotov \etal\ (1998) find 41 WR galaxies, defined
by broad emission between $\sim$ 4620 -- 4700 \AA. Often several broad
features are pointed out in the blue bump (\heii, \niii, but also other
lines they identify as N\,{\sc iii} $\lambda$ 4510, N\,{\sc ii} $\lambda$ 
4565, N\,{\sc v} $\lambda$4605,4620, C\,{\sc iv}$\lambda$4658).
According to their study 28 spectra also show broad \civ. 
Finally, few detections of broad \heiib, 
C\,{\sc iii}$\lambda$5696, and also C\,{\sc ii}$\lambda$4267 are signaled.
We retain all except one WR galaxy (Mrk 1026=NGC 848 showing no broad \heii)
from their study.

\subsection{Narrow-band imaging and others}
Drissen \etal\ (1993) have conducted a search for \heii\ emission
via narrow-band imagery in four low mass galaxies (GR8, NGC 2366,
IC 2574, NGC 1569). Two of them are now confirmed WR galaxies (see above),
IC 2574 remains to be studied spectroscopically, and GR8 yielded
negative results (no \Heii).

Schmidt-Kaler \& Feitzinger (1984) initiated a search for 30 Dor
and NGC 604 like objects based on POSS, ESO-Blue, and SRC film.
To the best of our knowledge results from this survey have not been
published.

%%%%%%%%%%%%%%%%%%%%%%%%%%%%%%%%%%%%%%%%%%%%%%%%%%%%%%%%%%%%%%%%%%%%%%%%
\section{New catalogue of WR galaxies}
\label{s_catalogue}
In this Section we present an updated list of all ``WR galaxies'' outside
the Local Group.
As pointed out earlier we remind the reader that the working definition
for WR galaxies refers only to the detection of a broad WR feature in 
an integrated spectrum, irrespectively of the included area.
This fact must be taken into account in subsequent interpretations.
By this working definition we also include some nearby objects 
(e.g.\ NGC 300) which, although already known to harbour WR stars, were 
not included in the catalogue of Conti (1991).
An inventory of the Galactic WR stars is found in the 7th catalogue of 
van der Hucht \etal\ (1998; cf.\ van der Hucht 1996).
For individual WR stars and WR populations in the Local Group see
e.g.\ the review of Massey (1996).
% Lequex \& Azzopardi (1991) --> old review on LG...

\subsection{Selection criteria and procedure}
During the last three years the literature was followed for any 
publication relating to possible or confirmed signatures from WR stars
in extragalactic objects. At the end of july 1998 we also made a 
systematic search in the Astrophysics Data System (ADS) for the 
occurrence of ``WR'', ``W-R'', ``Wolf-Rayet'', or ``Wolf Rayet'' in all 
the included literature starting with 1970. The search yielded 1780 
references to refereed papers, and 1061 other references, all of which 
were carefully inspected. In addition all publications from the IAU 
Symposia 67, 80, 83, 88, 99, 105, 116, 122, 143, 149, 163 were scanned.
Although a complete inspection is not possible the present procedure
should guarantee a fairly high degree of completeness.

For the inclusion in our list of ``WR galaxies'' the criteria
are {\em 1)} the presence of {\em broad} emission at 4686 \AA\ due to
\Heii, or {\em 2)} a broad ``WR bump'' at $\sim$ 4660 \AA, or 
{\em 3)} other broad emission lines attributed to WR stars.
The second criterion accounts for the difficulty of resolving
\heii\ from other emission lines in medium resolution spectra
(see e.g.\ Kunth \& Joubert 1985), but may in some cases introduce objects
where nebular emission lines dominate the WR bump. Implicitly
both criteria {\em 1)} and {\em 2)} were adopted in the earlier compilation
of Conti (1991).
The third criterion allows for objects where broad carbon lines
(e.g.\ \ciiib, \civ) presumably from WR stars are detected.
In practice only one case (NGC 1365 for which no blue spectrum was
available) falls in this category.

The presence/absence of features and their qualification (broad and/or nebular)
is generally not fully objective (see Sect.\ \ref{s_search} for more
details). Furthermore in many cases the spectra are not available for 
inspection. 
We therefore follow the procedure or judgement of the authors of the 
original publication.

All objects resulting from this selection are listed in Table
\ref{ta_wrgals}. Obviously most of the objects appear in various
catalogues. We have chose the following labeling priority:
first the Messier number, then the NGC, the Markarian, the
Zwicky lists, or finally other labels (UM, Tol, SBS, etc).
Often several regions in the same object show WR features. In this case
only the main object name is listed (column 1); information about the 
different regions are given in Sect.\ \ref{s_objects}.

The general properties of objects, 
equatorial coordinates (equinox 1950, cols.\ 2 and 3), 
morphological type (col.\ 4), 
apparent blue magnitude (col.\ 5), 
and heliocentric radial velocity (col.\ 6) 
have been extracted from NED. 
Column 7 gives the reference of the first publication indicating
the presence of {\em broad} \heii. ``(C91)'' indicates that this objects
was already included in the compilation of Conti (1991).
Col.\ 8 gives information about the existence of {\em nebular}
\heii\ (see footnote of the Table for the keys).
The source of this information is found under the remarks on
each individual object (Sect.\ \ref{s_objects}).
Col.\ 9 gives the reference to the first detection of broad \civ\
(exceptionally also other broad features, cf.\ NGC 1365).

%dsnumber next 5 lines
Our current compilation (Table \ref{ta_wrgals}) includes 139 WR galaxies
(cf.\ 37 WR galaxies in Conti 1991).
Among them 57 objects show also broad \civ\ features. In addition to the
broad (stellar) \heii\ emission, a nebular \Heii\ component is well established
(suspected) in 44 (54) objects.

The database will be made available electronically at the CDS and through 
the Web\footnote{informations at \\
{\tt http://www.obs-mip.fr/omp/astro/people/schaerer}}.
It is our intention to update the catalogue in the future by this means.

%%%%%%%%%%%%%%%%%%%%%%%%%%%%%%%%%%%%%%%%%%%%%%%%%%%%%%%%%%%%%%%%%%%%%%%%
% Maffei 1, Sculptor, M81, M83 groups
%\section{WR stars in groups surrounding the LG}
% --> included below...

%%%%%%%%%%%%%%%%%%%%%%%%%%%%%%%%%%%%%%%%%%%%%%%%%%%%%%%%%%%%%%%%%%%%%%%%
% outside LG and surrounding groups -- ``REAL'' WR-galaxies
\section{Remarks on individual WR galaxies}
\label{s_objects}
Below a brief comment is given for each object from Table \ref{ta_wrgals}
on the detected WR features and possible nebular \heii.
In some objects the strongest nebular lines also show broad and/or
asymmetric components, attributed to gaz-flows or broad stellar 
emission. These finding are also reported below.

{\em NGC 53         } ---
Detection of \heii\ by Masegosa \etal\ (1991) from automatic search in 
\hii\ galaxy catalogue of Terlevich \etal\ (1991).

{\em NGC 55         } ---
Rosa \& D'odorico (1986) surveyed eight giant \hii\ regions in NGC 55.
Two were found with broad \heii\ and \civ\ features.

{\em UM 48          } ---
Automatic detection by Masegosa \etal\ (1991).
WR features unlikely according to the analysis of Pindao \etal\ (1999).

{\em Mrk 960        } ---
Broad features around 4700 \AA\ were first suspected by S.\ Consid\`ere (private 
communication). A better spectrum of this galaxy confirms the presence of a broad 
\heii\ emission line (Kovo \& Contini 1998).

{\em NGC 300        } ---
Fifteen positions were observed by D'Odo\-ri\-co \etal\ (1983) in this Sculptor
group galaxy often considered as a twin of M 33. Two regions showed \heii\, 
one also \civ.
Several investigations were undertaken subsequently to search for individual
WR stars or small clusters. The latest work (Breysacher \etal\ 1997, cf.\ 
references therein) detected 12 WR stars increasing the total to 34 known
WR stars.

{\em IRAS 01003-2238} ---
Broad \heii\ and \niii\ emission has been observed by Armus \etal\ (1988).
Other broad features are tentatively detected. Based on the 
equivalent withs of the WR-bump they estimate $\sim$ 10$^5$ WR stars
are present in this luminous infrared galaxy, the most distant WR galaxy known 
so far. 

{\em UM 311         } ---
This \hii\ region is possibly located in the galaxy NGC 450.
Automatic detection of broad \heii\ by Masegosa \etal\ (1991).
The high S/N spectrum of Izotov \& Thuan (1998) shows 
both broad \heii\ and \civ\ features, but no nebular \Heii.
A nebular component is also detected by \git.
Pindao (1998) confirms the presence of a broad WR bump.

{\em Tol 0121-376   } ---
Masegosa et al.\ (1991) and Pindao (1998) signal the presence of a broad
WR bump in this galaxy.

{\em Minkowski's Object   } ---
A weak broad \heii\ feature and possible nebular \Heii\ has been found 
by Breugel \etal\ (1985) in this ``starburst triggered by a radio jet''.
They also point out a close similarity of the emission line spectrum with 
the NGC 7714, a prototypical starburst galaxy.

{\em Mrk 996        } ---
This unusual blue compact galaxy shows \heii, \niii\ and \civ\ features
in the HST FOS spectra of Thuan \etal\ (1996).

{\em Mrk 589        } ---
\git\ point out the presence of broad \heii\ and several N lines
in the blue bump, as well as broad \civ. The S/N in the red appears quite
low for its detection.

{\em UM 420         } ---
Broad \heii\ emission has been found by Izotov \& Thuan (1998).
From their reanalysis \git\ signal also nebular \Heii\ and broad \civ.
The S/N appears fairly low for the latter assertion.
%They also note a high frequency ($\sim$ 2/3) of WR detections in 
%their low metallicity \hii\ region sample.

{\em Mrk 1039       } ---
The presence of nebular and broad \heii\ and \civ\ has been
discovered by Huang \etal\ (1998) in this \hii\ galaxy.

{\em Tol 0226-390   } ---
Masegosa et al.\ (1991) and Pindao (1998) signal the presence of a broad
WR bump in this galaxy.

{\em Tol 0242-387   } ---
Masegosa et al.\ (1991) and Pindao (1998) signal the presence of a broad
WR bump in this galaxy.

{\em Mrk 598        } ---
Pindao (1998) signals the presence of a broad WR bump in the western knot of
this galaxy.

{\em NGC 1140       } ---
A broad WR bump and \civ\ are detected by \git.

{\em NGC 1156       } --- 
Emission features of WR stars and a high-excitation \hii\ region
in the nucleus are signaled by Ho \etal\ (1995) from their 
magnitude limited survey of nuclei of nearby galaxies.
A close resemblance of the spectrum with NGC 4214 is pointed out.
They also signal a broad \halpha\ component.

{\em NGC 1313       } ---
WR features are found in two regions at large galactocentric radii
of this Transition Magellanic galaxy (Walsh \& Roy 1997).
For region \# 28 both \heii\ and \civ\ are found. No detailed 
information is provided  about region \# 3.
Pindao \etal\ (1999) also signal the possible detection of broad \heii;
exact position unknown.

{\em NGC 1365       } ---
The detection of broad \ciiib\ and \civ\ (marginally) in this giant
extragalactic \hii\ region was made by Phillips \& Conti (1992).

{\em SBS  0335-052  } ---
The first detection of nebular \heii\ in this very low metallicity
object was reported by Izotov \etal\ (1990).
Nebular and broad \heii\ have been found in the reanalysis 
of Izotov \etal\ (1998); no broad features had been signaled by Izotov \etal\ (1997c).

{\em NGC 1510       } ---
Eichendorf \& Nieto (1984) show the presence of broad \heii\ in one
component of this amophous galaxy. See the discussion in Conti (1991) 
for more details.

{\em NGC 1569       } ---
Narrow-band $\lambda$4686 filter imaging of Drissen \etal\ (1993) 
revealed the possible presence of WR stars.
Spectroscopy by Drissen \& Roy (1994) in the outskirts of the galaxy
shows broad \heic\ and He\,{\sc i} $\lambda$6678 both attributed to a 
late WN star.
According to  Gonzal\'ez-Delgado \etal\ (1997) this region is located 
4\arcsec\ west of super star cluster (SSC) {\em A}.
Ho \etal\ (1995) find the WR bump and a broad \halpha\ component
``in the nuclear spectrum''.
Gonzal\'ez-Delgado \etal\ (1997) find several broad features in the WR 
bump which are confined to SSC {\em A}.
Broad \heii\ has also been detected by Martin \& Kennicutt (1997).

{\em NGC 1614       } ---
A broad WR bump is detected by Pindao (1998). The measurement of 
Vacca \& Conti (1992) provides an upper limit on broad \heii.

{\em VII Zw 19      } ---
Found by Kunth \& Joubert (1985) in their ``survey'' of 45 blue emission
line galaxies for showing excess emission between 4600 and 4711 \AA.

%{\em Mrk 1087       } ---
%Broad \heii\ was detected by Vacca \& Conti (1992).
%ds only UPPER LIMIT GIVEN !

{\em NGC 1741       } ---
This well studied galaxy is part of the Hickson (1982) ``compact'' group \# 31
and interacting with H31 (see below).
Several broad components in the WR bump (\heii, \niii) and nebular emission
of [Fe\,{\sc iii}] $\lambda$4658 and \heii\ have been identified 
by Kunth \& Schild (1986).
According to Conti (1991) \niii\ cannot be confirmed.
The spectrum of \vc92\ of region B is shown in Conti \etal\ (1996)
who obtained also a UV spectrum with GHRS on HST of this region.
An upper limit on \heii\ is also given by  Vacca \& Conti (1992) for
region A.
The high S/N spectrum of Izotov \& Thuan (1998) shows broad \heii,
whereas the \civ\ is absent. 
%NO Nebular according to Izotov: oct 26, 98. Error in Tables...
No nebular \Heii\ component is present according to the reanalysis of \git.

{\em H31A           } ---
A broad \heii\ feature and was pointed out by Rubin \etal\ (1990) in galaxy
A and possibly also in H31C = NGC 1741 (see above).

{\em Mrk 1094       } ---
\kj85\ list this object as having a broad \heii\ excess above 0.8 $\sigma$ of
the background.
Broad \heii\ was detected by Vacca \& Conti (1992) in region A. 
Upper limits on \Heii\ are given for two other regions.

{\em II Zw 40       } ---
Broad \heii\ was detected by Kunth \& Sargent (1981) in this well studied
low metallicity galaxy.
See also \vc92\ for more recent observations and a study of its WR content.
Broad \heii\ and possibly also \civ\ are signaled by Martin (1997).
An important contamination by nebular \heii\ has been suspected by Schaerer
(1996) from its similarity with Pox 4 and on theoretical grounds.
A broad asymmetric emission components of \halpha\ has been found by 
M\'endez \& Esteban (1997).
\git\ find the presence of both stellar and nebular \Heii\ and broad
\civ.

{\em Tol 0633-415   } ---
Automatic detection by Masegosa \etal\ (1991). Pindao \etal\ (1999)
classify this as a suspected WR galaxy.

{\em Mrk 5          } ---
The high S/N spectrum of Izotov \& Thuan (1998) shows broad \heii.
No nebular \Heii\ component is present according to the reanalysis of \git.
%NO Nebular according to Izotov: oct 26, 98. Error in Tables of IT98.

{\em IRAS 07164+5301} ---
Huang \etal\ (1996) detect the presence of broad lines around 4686 \AA\
suggesting \niii, \ciii, and \heii\ in this IRAS source. 
They also indicate a tentative detection of O\,{\sc v} $\lambda$5835 
and a lack of \civ. The spectrum is of fairly low S/N.

{\em Mrk 1199       } ---
Izotov \& Thuan (1998) signal nebular and broad \heii\ and \civ\ features
from their high S/N spectrum\footnote{The reported nebular \heii\ intensity in
Izotov \& Thuan (1998) is erroneous according to Y.\ Itozov (1998, private
commnication). No indication is therefore given in column 8 of table \ref{ta_wrgals}
for this object.}. The latter features appears quite weakly.
No nebular \Heii\ component is present according to the reanalysis of \git.

{\em NGC 2363       } ---
This is a well studied giant \hii\ region consisting of two main knots
and located south west of the irregular 
galaxy NGC 2366 (e.g.\ Drissen \etal\ 1993, Gonzal\'ez-Delgado \etal\ 1994,
and references therein). 
Spectra of the region taken up to 1992 all show narrow \heii\ (see references
in Drissen \etal\ 1993). From narrow-band imaging Drissen \etal\ (1993) find excess
\heii\ emission in both knots, but much stronger in the fainter eastern knot.
They argue for WR stars in this knot (B).
Spectroscopy by Gonzal\'ez-Delgado \etal\ (1994) confirms the presence
broad and nebular \heii\ in knot B, and detect also \civ\ attributed to 
WC stars.
\yu97\ find broad and nebular \heii\ emission in knot A, and nebular \Heii\
in knot B (cf.\ also the reanalysis by \git).
Broad emission component of \halpha, \hbeta, and [O\,{\sc iii}] are also
known (Roy \etal\ 1992, Izotov \etal\ 1997a).

{\em Mrk 8          } ---
\kj85\ list this object as having a broad \heii\ excess above 0.8 $\sigma$ of
the background.

{\em NGC 2403       } ---
Drissen \& Roy (1996) detect broad \heii\ and \civ\ in two giant \hii\ regions in 
this galaxy of the M81 group.

{\em VII Zw 187     } ---
\kj85\ list this object as having a broad \heii\ excess above 0.8 $\sigma$ of
the background.

{\em SBS 0749+582   } ---
Broad and nebular \heii\ is signaled by \yu97.

{\em Mrk 1210       } ---
Storchi-Bergmann \etal\ (1998) detect the presence of a broad \heii\ 
component attributed to WR stars in the nucleus of this Seyfert 2 galaxy.
A broad WR bump is also detected by Pindao (1998).

%{\em UGC 4305       } ---
%The presence of WR stars in this irregular galaxy has been inferred 
%from IUE spectra by Lamb \etal\ (1990).

{\em IRAS 08208+2816} ---
The presence of nebular and broad \heii\ and broad \civ\ was found
by Huang \etal\ (1998) in this luminous infrared galaxy.
The WR bump luminosity is exceptionally large and rivals
that of IRAS 01003-2238. Interestingly the authors also find
essentially zero internal reddening derived from the Balmer
decrement.

{\em He 2-10        } ---
This dwarf emission galaxy can be considered the ``prototypical
WR galaxy'' since it was the first where \heii\ emission
attributed to WR stars was detected (Allen \etal\ 1976).
Abundant observational data is available for this galaxy.
The detection of Allen \etal\ has been confirmed by Hutsemekers 
\& Surdey (1984), who also suspected \civ\ emission due to WC stars
from their spectrum.
Broad \heii\ and a weak \civ\ feature were detected by 
Vacca \& Conti (1992) in their region {\em A}. 
Both features were confirmed by Schaerer \etal\ (1999)
from their high S/N spectra.
An upper limit for \heii\ in region B is also given in \vc92.
The HST UV images of Conti \& Vacca (1994) resolve this galaxy in
multiple knots.
Broad asymmetric emission components of \halpha\ and 
[N\,{\sc ii}] $\lambda$6584 have been found by M\'endez \& Esteban (1997).

{\em Mrk 702        } ---
An incorrect object name (C 0842+163) was used for this galaxy by 
Masegosa \etal\ (1991), who report a broad WR feature.
\git\ detect both broad \heii\ and \civ\ features.
The S/N appears fairly low for the latter assertion.

%{\em Zw 0855+06     } ---
%Broad \heii\ was detected by Vacca \& Conti (1992).
%ds only UPPER LIMIT GIVEN !

{\em SBS 0907+543   } ---
Broad and nebular \heii\ has been found by \yu97.

{\em SBS 0926+606   } ---
Broad and nebular \heii\ has been found by \yu97\ (cf.\ also \git).
They also indicate the presence of low intensity broad components
of \halpha\ and/or \oiiib.

{\em I Zw 18        } ---
This well-known object is the galaxy with the lowest metal content
known. While nebular \heii\ was observed for a long time, only recently
the deep spectra of Izotov \etal\ (1997b) and Legrand \etal\ (1997)
revealed several broad emission components (\ciii, \heii, \civ) attributed 
to WN and WC stars. The spatial distribution of \heii\ emission
was studied by Hunter \& Thronson (1995) and De Mello \etal\ (1998)
based on WFPC2 HST observations.
\yu97\ indicate the presence of low intensity broad components
of \halpha\ and/or \oiiib.

{\em ESO 566-7      } ---
An incorrect object name C 0942-1929A was used by Masegosa \etal\ (1991) 
for this galaxy, where they report a broad WR feature. 
Pindao (1998) confirms the presence of the WR bump.

{\em NGC 3003       } ---
A complex broad WR-bump was signaled by Ho \etal\ (1995),
who also note the absence of a broad \halpha\ component in contrast
to their spectra of other WR galaxies (cf.\ NGC 1156, NGC 1569, NGC 4214).

{\em Mrk 22         } ---
Broad and nebular \heii\ has been found by \y94. Their spectrum appears 
noisy to detect \civ. Its detection is, however, signaled by \git.

{\em Mrk 1236       } ---
Kunth \& Schild (1986) point out the presence of broad \heii\ in this 
galaxy. Broad \heii\ was detected in region A by \vc92.
Observations of \git\ signal the presence of nebular and broad \Heii, 
broad \civ\ and possibly also C\,{\sc iii} $\lambda$5696. 
A broad emission feature is identified as C\,{\sc ii} $\lambda$ 4267.

{\em SBS 0948+532    } ---
Broad and nebular \heii\ has been found by \y94.
Also \civ\ is present according to the reanalysis of \git,
although the S/N appears fairly low.

{\em NGC 3049       } ---
This Virgo Cluster galaxy shows broad \nii\ and \heii\ (Kunth \& Schild
1986). The WR bump is confirmed by Mas-Hesse \& Kunth (1991, 1998),
Masegosa \etal\ (1991), and Pindao (1998).
Broad \heii\ was detected in region {\em A} by \vc92; an upper limit is given
for region {\em B}.
The high S/N observations of Schaerer \etal\ (1999) reveal broad features
of \niii, \heii, \ciiib, and \civ\ testifying of the presence of late-type WN
and late-type WC stars.
These features are confirmed by \git.

{\em Mrk 712        } ---
Contini \etal\ (1995) and Contini (1996) signal the presence of broad 
\niii\ and \heii\ in a giant \hii\ region of this IRAS barred spiral galaxy.
A nebular contribution to \Heii\ may be present (Contini 1996).

{\em Tol 0957-278   } ---
\kj85\ list this object (=Tol 2) as having a broad \heii\ excess above 0.8 $\sigma$ of
the background in the NE component.
Upper limits of \heii\ are given for 2 regions by Vacca \& Conti (1992).
Possible detection of broad \heii\ according to Pindao \etal\ (1999).

{\em NGC 3125       } ---
Broad \heii\ emission was found by Kunth \& Sargent (1981; cf.\ also Kunth
\& Joubert 1985) in this dwarf galaxy. They also note narrow \feiiia\ emission.
Broad \heii\ was detected in two regions by Masegosa \etal\ (1991) and \vc92.
Pindao (1998) reconfirms the detections of Masegosa \etal.
The high S/N observations of Schaerer \etal\ (1999) reveal broad features
of \niii, \heii, and \civ\ in both regions, testifying of the presence of late WN
and early WC stars. 

{\em Tol 1025-284   } ---
Pindao (1998) signals the presence of a broad WR bump in this galaxy.

{\em Mrk 33         } ---
\kj85\ list this object as having a broad \heii\ excess above 0.8 $\sigma$ of
the background.
The WR bump is confirmed by the spectra of Mas-Hesse \& Kunth (1991, 1998).

{\em Mrk 178        } ---
The SE knot of this galaxy shows broad \heiia\ according to Gonz\'alez-Riestra
\etal\ (1984). \Heii\ emission was already noted by Sargent (1972).
\git\ signal the presence of nebular and broad \Heii, and \civ.

{\em Mrk 1434       } ---
Broad and nebular \heii\ has been found by \yu97.
They also indicate the presence of low intensity broad components
of \halpha\ and/or \oiiib.
The reanalysis of \git\ also shows \civ\ emission.

{\em Mrk 1259       } ---
Ohyama \etal\ (1997) detect broad \niii\, and \heii\ lines in the nuclear
spectrum of this nearby starburst galaxy. Their study suggests the existence
of a superwind seen nearly pole-on (see also Ohyama \& Taniguchi 1998).
\git\ signal also the presence of \civ\ and no nebular \Heii.
A broad feature identified as N\,{\sc iii} $\lambda$4510 is also indicated.

{\em Mrk 724        } ---
Kunth \& Schild (1984) find broad \heii\ and a broad feature close to \civ, 
as well as additional nebular lines ``contaminating'' the WR bump (\feiiia, 
\heii, \ariva).
The identification of \civ\ is not well established (see Kunth \& Schild, 1984; 
Conti, 1991).

{\em NGC 3353       } ---
Steel \etal\ (1996) report the presence of a broad WR bump and a possible
detection of \civ\ in region {\em A}.
These signatures are confirmed by Huang \etal\ (1998) who also find indications
for nebular \heii.

{\em NGC 3367       } ---
The presence of a broad WR bump in the LINER nucleus has been signaled 
by Ho \etal\ (1995), who also note unusually broad and asymmetric emission lines (e.g.\ \hbeta).

{\em NGC 3395       } --- 
Weistrop \etal\ (1998) signal the presence of broad \heii\ and \civ.

{\em NGC 3396       } --- 
Weistrop \etal\ (1998) signal the presence of broad \heii\ and \civ, as well 
as nebular \heii.

{\em Mrk 1271       } ---
The spectrum of Izotov \& Thuan (1998) shows weak nebular and broad \heii\
features. Although very weak, \civ\ could also be present according to \git.
Contini (1996) only finds nebular emission lines.

{\em SBS 1054+365   } ---
Broad \heii\ has been detected by \yu97.
This is confirmed by the reanalysis of \git.

{\em Mrk 36         } ---
The spectrum of Izotov \& Thuan (1998) shows nebular and broad \heii\
features. No WR signature was detected by Campbell \etal\ (1986) and 
Mas-Hesse \& Kunth (1991, 1998).

{\em NGC 3690       } ---
Ho \etal\ (1995) signaled the presence of WR features in several
regions of this galaxy (=Arp 299); they exclude WR features in the 
nucleus. Vacca (1996, private communication) signals the presence of broad 
\heii\ in Arp 299B and Arp 299C which are part of this complex
system.

%{\em Mrk 178        } ---
%Broad and nebular \heii\ and very strong \civ\ are detected by \git.

{\em NGC 3738       } ---
Martin (1997) points out the presence of broad \heii.

{\em UM 439         } ---
Automatic detection by Masegosa \etal\ (1991).
Also listed as WR candidate by Pindao \etal\ (1999).

{\em Mrk 182        } ---
Broad \heii\ is pointed out by \git. The spectrum may be too 
noisy to establish the presence/absence of a nebular component.

{\em Mrk 1450       } ---
Nebular and broad \heii, and \civ\ have been found by \y94\ (cf.\
also \git).

{\em Mrk 1304       } ---
Automatic detection of \Heii\ by Masegosa \etal\ (1991),
later confirmed by Pindao (1998).
Nebular and broad \heii, and \civ\ have been found by \git.
The S/N appears fairly low for the latter assertion.

{\em Mrk 1305       } ---
Broad \heii\ and \civ\ have been found by \git.
The S/N appears fairly low for the latter assertion.

%{\em Arp 248b       } ---
%The presence of WR stars in this merging galaxy has been inferred 
%from IUE spectra by Lamb \etal\ (1990).

{\em Mrk 750        } ---
\heii\ emission was first signaled by Kunth \& Joubert (1985).
According to Conti (1991) \niii\ is also detected in a spectrum
of Salzer.
These detections are consistent with the spectrum of 
Izotov \& Thuan (1998) showing a broad WR bump and nebular \Heii.
The reanalysis of \git\ also reveals \civ\ emission.

{\em Pox 4          } ---
\kj85\ list this object as having a broad \heii\ excess above 0.8 $\sigma$ of
the background.
This object is also included in the studies of Campbell \etal\ (1986) and 
Masegosa \etal\ (1991)
\footnote{Campbell \etal\ refer to C 1148-203. In Masegosa \etal\ the name 
C 1148-2020 (see their Table 1) and erroneously Tol 1148-202 (Table 2) is used. 
The proper identification of this Cambridge object (see Telles \etal\ 1997) 
is IRAS 11485-2018 = Pox 4 according to NED.},
which signal a possible WR feature in one or two regions.
The reanalysis of Pindao (1998) confirms the WR bump in one region.
Broad \heii\ has been measured by Vacca \& Conti (1992) in two regions.
The existence of a broad component is not well established (e.g.\ Kunth 
\& Sargent, 1981).
Nebular \heii\ is most likely present in region {\em A} (see spectrum of
Vacca \& Conti 1992).
A broad asymmetric emission component of \oiiib\ has been found by 
M\'endez \& Esteban (1997).

{\em UM 461         } ---
Conti (1991) reports the presence of broad \heii\ and relatively
strong nebular [Ar\,{\sc iv}] lines in a spectrum from Salzer.

%{\em UM 462         } ---
{\em Mrk 1307        } ---
Izotov \& Thuan (1998) indicate the presence of broad \heii\ and
nebular \Heii\ (cf.\ Guseva \etal\ 1998). 
The presence of a broad component appears somewhat
marginal (see also Contini 1996).

{\em Mrk 193         } ---
The analysis of \git\ indicates the presence of broad and nebular \heii\ 
in this object; the broad component was not signaled by \y94.

{\em ISZ 59         } ---
\kj85\ list this object as having a broad \heii\ excess above 0.8 $\sigma$ of
the background.

{\em NGC 3995       } ---  
Weistrop \etal\ (1998) signal the presence of broad \heii\
and \civ, as well as nebular \heii.

{\em NGC 4038       } ---
From ten giant \hii\ regions surveyed by Rosa \& D'Odorico (1986) in this galaxy
of the Antennae, one exhibits a broad emission feature at the blue WR bump.
 
{\em SBS 1211+540   } ---
The reanalysis of \git\ shows nebular and broad \heii; \y94 only signaled
nebular emission.

{\em NGC 4214       } ---
WR signatures in this galaxy were found independently by Mas-Hesse \& Kunth (1991;
cf.\ also 1998) and Sargent \& Fillippenko (1991).
The observations of the former show a broad WR bump around 4650 \AA\ and 
\civ\ due to WC stars. The latter detect WR signatures of \niii, \ciii, possibly
also \ciiia, and \heii\ in two knots. From the C lines they also suspected the
presence of WC stars.
Broad \heii\ and \civ\ in several regions has also been signaled by Martin \& 
Kennicutt (1997).
A broad \halpha\ component, attributed to WN stars, was also detected by
Sargent \& Fillippenko in one knot.
Recent UV spectroscopy of NGC 4214 with HST was obtained by Leitherer 
\etal\ (1996).
Detailed spectroscopic spatial mapping by Ma\'{\i}z-Apell\'aniz et
al.\ (1998) shows the presence of \heii\ (broad and narrow) in several
regions of NGC 4214.

{\em NGC 4216       } ---
From five giant \hii\ regions surveyed by Rosa \& D'Odorico (1986) in this galaxy
two exhibit a broad emission feature at the blue WR bump.

{\em NGC 4236       } ---
Gonz\'alez-Delgado \& Perez (1994) report the presence of broad \niii\ and
\heii\ features in their \hii\ region III.

{\em M 106          } ---
Castellanos \etal\ (1998) point out the presence of broad \heii\ in the brightest
\hii\ region of this Seyfert 2 galaxy.

{\em SBS 1222+614   } ---
Nebular and broad \heii, and \civ\ have been found by \yu97\ (cf.\ Guseva \etal\ 1998).

{\em NGC 4385       } ---
WR signatures of \niii, \heii\ and possibly also \civ\ have been detected by
Campbell \& Smith (1986).
The first two lines are also found in the spectrum of Durret \& Tarrab (1988), 
who signal also a possible detection of \ciiia.
The findings are confirmed by Conti (1991) according to a spectrum of Salzer;
according to Conti \ciiia\ has more likely to be identified with \feiiia.

{\em II Zw 62       } ---
\kj85\ list this object as having a broad \heii\ excess above 0.8 $\sigma$ of
the background.

{\em Mrk 209        } ---
Nebular and broad \heii\ have been found by \yu97; the reanalysis of \git\ 
also reveals \civ\ emission, although very weak.

{\em NGC 4449       } ---
Martin \& Kennicutt (1997) indicate the presence of broad \heii\ and \civ\
in several regions of this object.

{\em NGC 4532       } ---
The presence of a weak broad WR bump has been signaled by Ho \etal\ (1995),
who also note a very weak broad \halpha\ component.

{\em Mrk 1329       } ---
Broad \heii\ and a possible detection of \civ\ are signaled by \git.
He\,{\sc ii} $\lambda$5412 emission is also pointed out.

{\em Tol 1235-350   } ---
Pindao (1998) signals the presence of a broad WR bump in this galaxy.

{\em NGC 4670       } ---
Mas-Hesse \& Kunth (1991, 1998) report the presence of a WR bump in this
galaxy; the latter publication provides only an upper limit.

{\em Tol 1247-232   } ---
Automatic detection of broad \heii\ by Masegosa \etal\ (1991) confirmed
by the analysis of Pindao (1998).

{\em SBS 1249+493   } ---
The reanalysis of \git\ indicates the presence of nebular and broad \heii; 
previously only a narrow component was detected (Thuan \etal\ 1995).

{\em NGC 4861       } ---
The spectrum of Dinerstein \& Shields (1986) shows a broad WR feature
centered at 4686 \AA\ and a possible detection of \civ\ (but cf.\ Conti 1991).
The WR bump is confirmed by Mas-Hesse \& Kunth (1991, 1998), 
Motch \etal\ (1994) who also find nebular \Heii, and by
Martin \& Kennicutt (1997) who possibly also find \civ.
The spectrum of \yu97\ shows broad and nebular \Heii, as well as 
\civ\ (cf.\ Guseva \etal\ 1998).

{\em Tol 30         } ---
Broad \niii, \heii, and probably also nebular \Heii\ have been found by
Contini (1996). The exact region is not specified.
Pindao \etal\ (1999) signal the possible presence of the WR bump
in the \hii\ region Tol 1303-281 NW associated with this galaxy.

{\em Pox 120        } ---
\kj85\ list this object as having a broad \heii\ excess above 0.8 $\sigma$ of
the background.

{\em Pox 139        } ---
\kj85\ list this object as having a broad \heii\ excess above 0.8 $\sigma$ of
the background.
This finding is confirmed by the  observations of \vc92.

{\em NGC 5068       } ---
From five giant \hii\ regions surveyed by D'Odorico \etal\ (1983) in this galaxy,
two exhibit a broad emission feature at the blue WR bump.

{\em SBS 1319+579   } ---
Nebular and broad \heii\ have been found by \yu97
In their reanalysis \git\ also detect \civ.
%*** CHECK if 3 different objects as listed in \yu97 !?
% NO: only one spectrum: probably sum !

{\em NGC 5128       } ---
The observations of six \hii\ regions in the elliptical galaxy Cen A 
(classified as Seyfert 2 in NED) by 
M\"ollenhoff (1981) revealed several WR features in one region (\# 13) 
near the rim of the dust disk of this well studied galaxy.
Broad lines of \niii, \heii, \ciii, and \civ\ are identified.
According to Rosa \& D'Odorico (1986), WR features are found in two
(including \# 13) out of six surveyed regions.

{\em Pox 186        } ---
\kj85\ list this object as having a broad \heii\ excess above 0.8 $\sigma$ of
the background.

{\em Tol 35         } ---
\kj85\ list this object as having a broad \heii\ excess above 0.8 $\sigma$ of
the background.
Campbell \etal\ (1986) indicate only nebular \Heii.
The WR feature is confirmed by the spectrum of Campbell \& Smith (1986),
by Masegosa \etal\ (1991), \vc92, and Pindao (1998).
Broad asymmetric emission components of \halpha\ and \oiiib\ have been
found by M\'endez \& Esteban (1997).

{\em M 83           } ---
From eight giant \hii\ regions surveyed by Rosa \& D'Odorico (1986) in this galaxy,
three exhibit a broad emission feature at the blue WR bump.

{\em NGC 5253       } ---
This is a well studied amorphous galaxy.
WR features have not been detected by Rosa \& D'Odorico (1986); four regions
surveyed.
The first reports of a broad WR bump in this galaxy are from Campbell \etal\
(1986) and Walsh \& Roy (1987).
The high S/N observations of Schaerer \etal\ (1997) reveal broad features
of \niii, \heii, and \civ\ in two regions, testifying of the presence of late-type WN
and early-type WC stars. Nebular \heii\ is very likely present in their region {\em A};
a region of exceptionally strong \heii\ (probably nebular) was also found
(see Schaerer \etal\ 1999).
%
%Martin \& Kennicutt (1997) report the presence of broad \heii\ and possibly
%also \civ.
The WR bump was also detected by Kobulnicky \etal\ (1997), Martin \& Kennicutt 
(1997), Mas-Hesse \& Kunth (1998) and Pindao (1998).

{\em Mrk 67         } ---
Conti (1991) reports a possible detection of a weak and broad \heii\ emission
feature from a spectrum of Salzer.

{\em Mrk 1486       } ---
Nebular and broad \heii\ have been found by \yu97.

{\em Tol 89         } ---
Durret \etal\ (1985) report the detection of a broad WR bump (likely \niii\ and
\heii) in this giant \hii\ region of NGC 5398. IUE spectra also indicate 
the presence of WR stars (see Durret \etal\ 1985, Conti 1991), possibly also of the 
WC type.
Pindao (1998) also signals the presence of the WR bump.
The high S/N observations of Schaerer \etal\ (1999) reveal broad features
of \niii, \heii, and exceptionally strong \civ, testifying of the presence of 
late-type WN and early-type WC stars. 
These authors also suspect a contribution from nebular \heii.

{\em NGC 5430       } ---
Keel (1982) reports a broad WR bump (likely \niii\ and \heii) in the spectrum
of a bright \hii\ region, SE of the nucleus of this barred galaxy.

{\em NGC 5408       } ---
WR features have not been detected by Rosa \& D'Odorico (1986);
two regions surveyed.
Masegosa \etal\ (1991) signal a possible detection of broad \heii\
in their region {\em B}.
According to Motch \etal\ (1994) the \Heii\ emission is only nebular
and is mostly found in two regions (their \# 3 and 4). 
Kovo \& Contini (1998) report both broad and nebular \heii\ in two \hii\ 
regions of this galaxy with a faint broad \niii\ in one of them. 
The nebular lines of \feiiia\, \ariva, and \arivb\ are strong.

{\em NGC 5457 = M 101 } ---
Rosa \& D'Odorico (1986) report of seven surveyed \hii\ regions with five
detections of WR features from the spectra of D'Odorico \etal\ (1983).
According to the latter, however, only two regions, Hodge 40 and NGC 5461 
(see below), show clear broad \heii\ features; the remaining objects 
required confirmation with spectra at higher resolution.
Although uncertain, broad \heii\ emission in Hodge 40 was also pointed 
out by Rayo \etal\ (1982).

{\em NGC 5461       } ---
The presence of broad \niii\ and \heii\ in this giant \hii\ region of M101
was first pointed out by Rayo \etal\ (1982) and later confirmed by D'Odorico 
\etal\ (1983).

{\em NGC 5471       } ---
Rayo \etal\ (1982) find a weak \heii\ feature in this \hii\ region of M 101.
No information on its width is given.
D'Odorico \etal\ (1983) detect unbroadened lines identified as \ciiia\, \feiiia, 
\heii, He\,{\sc i} $+$ \ariva\, and \arivb, whose origin is likely
nebular.
Mas-Hesse \& Kunth (1991, 1998) mention the presence of a WR bump.

{\em SBS 1408+551A  } ---
Nebular and broad \heii, and \civ\ have been found by Izotov \etal\ (1996).

{\em CGCG 219-066   } ---
% = SBS 1415+437
Nebular and broad \heii\ emission is found according to Izotov \etal\ (1998); 
no broad features had been signaled by Thuan \etal\ (1995).

{\em Mrk 475        } ---
Conti (1991) reports a detection of a moderately strong \heii\ 
emission feature and possible \niii\ from a spectrum of Salzer.
The similarity with the spectrum of Mrk 750 is pointed out.
Nebular and broad \heii, and \civ\ have been found by \y94\ 
(cf.\ Guseva \etal\ 1998).
Their spectrum also show the presence of an unidentified broad
feature at $\sim$ 4200 \AA.

{\em Mrk 477        } ---
Heckman \etal\ (1997) detect the presence of a broad \heii\ component 
attributed to WR stars together with several other stellar signatures 
in the nucleus of this powerful Seyfert 2 galaxy.

{\em Tol 1457-262A} ---           % ESO 513-11 (A) 
A broad \heiia\ emission line has been reported by Contini (1996) in 
one of the members of the galaxy pair Tol 1457-262.
A broad WR bump is also detected by Pindao (1998). 
 
{\em Tol 1457-262B} ---          
Whereas Contini (1996) signals the presence of a purely nebular
\heii\ emission, a broad  WR bump is detected by Pindao (1998).

{\em SBS 1533+574B  } ---
Nebular and broad \heii, and also \civ\ have been found 
in the reanalysis of \git; no broad features had been signaled by \yu97.

{\em IC 4662        } ---
From two giant \hii\ regions surveyed by Rosa \& D'Odorico (1986) in this 
nearby galaxy, two exhibit a broad emission feature at the blue WR bump.
This is also confirmed by the study of Heydari-Malayeri \etal\ (1991).
Richter \& Rosa (1991) detect \heii\ and \civ\ in one cluster.
The WR bump was also detected by Mas-Hesse \& Kunth (1998).

{\em NGC 6500       } ---
Broad \heii\ may be tentatively detected in the LINER nucleus of this
galaxy (Barth \etal\ 1997).

{\em Fairall 44     } ---
Broad \heii\, \civ\ and \niii\ emission lines are detected in this dwarf 
galaxy (Kovo \& Contini 1998).

{\em NGC 6764       } ---
Osterbrock \& Cohen (1982) point out the presence of broad \niii\ and \heii\
in the spectrum of the nucleus of this barred spiral galaxy
also classified as LINER or Seyfert 2.
Conti (1991) argues that the line at 4660 \AA\ in their spectrum is likely \feiiia.
The spectrum of Eckart \etal\ (1996) confirms the broad features
of Osterbrock \& Cohen. They also signal excessive widths of \heic\ and
\halpha\ which they attribute to emission from WR stars. 
Broad \ciiib\ and \civ\ lines from WC stars are reported in the nucleus 
of this galaxy by Kunth \& Contini (1998).

{\em Tol 1924-416   } ---
Pindao (1998) signals the presence of a broad WR bump in this galaxy.
Kovo \& Contini (1998) indicate only nebular \heii.

{\em IC 4870        } ---
The presence of a broad WR bump is signalled by Joguet \& Kunth (1999) in this
galaxy classified as starburst or Seyfert 2. The spectral range at $\lambda >$ 
5300 \AA\ was not covered.

{\em IC 5154        } ---
%Joguet \& Kunth (1999) note the presence of a broad WR bump in one of the nuclei
%of this galaxy classified as starburst or Seyfert 2. No information available for
%$\lambda >$ 5300 \AA.
Joguet \& Kunth (1999) note the presence of broad \heii, \niii, and \civ\
in one of the two nuclei of this galaxy which they classify as
starburst although previously known as Seyfert 2.

{\em ESO 108-IG 01  } ---
The presence of broad \heii\ and \civ\ is signalled by Joguet \& Kunth (1999) in this
galaxy classified as starburst or Seyfert 2. 

{\em Mrk 309        } ---
Osterbrock \& Cohen (1982) point out the presence of broad \niii\ and \heii\ 
in this Seyfert 2 galaxy.
Conti (1991) argues that the line at 4660 \AA\ is likely \feiiia.
\civ\ and \ciiib\ emission, attributed to WC stars, is also tentatively
detected by Osterbrock \& Cohen.

{\em Mrk 315        } ---
\kj85\ list this object as having a broad \heii\ excess above 0.8 $\sigma$ of
the background.

{\em ESO 148-IG 02      } ---
WR features (\niii, \heii) in this infrared galaxy have been reported by 
Johansson \& Bergvall (1988).

{\em III Zw 107     } ---
\kj85\ list this object as having a broad \heii\ excess above 0.8 $\sigma$ of
the background in the southern component.

{\em Mrk 930        } ---
Broad \heii\ has been found by \yt98. The reanalysis  of \git\ also reveals
a nebular \Heii\ component and broad \civ.
The quality of the spectrum appears fairly low for the latter assertion.

{\em NGC 7714       } ---
Van Breugel \etal\ (1985) reported weak WR features near 4686 \AA\
and possible nebular \Heii\ emission
in the nucleus of this ``prototypical starburst'' galaxy, and call 
attention to the similarity with Minkowski's object (above) and 
extragalactic \hii\ regions.
New long-slit observations at several position angles were obtained
by Gonz\'alez-Delgado \etal\ (1995), confirming the presence of broad
\heii\ in the nucleus. Nebular emission can also be suspected from
their spectrum.
From the same observations Garc\'{\i}a-Vargas \etal\ (1997) find
broad WR bumps ($\sim$ 4660 \AA) in three extra-nuclear giant \hii\ regions.
One of them shows a definite detection of \civ; for the remaining
regions upper limits on \civ\ are given.
Pindao (1998) also signals the presence of a broad WR bump; 
no information about the observed location given.
       
%%%%%%%%%%%%%%%%%%%%%%%%%%%%%%%%%%%%%%%%%%%%%%%%%%%%%%%%%%%%%%%%%%%%%%%%
\section{\hii\ galaxies showing only nebular \Heii\ emission}
\label{s_heii}
\hii\ regions (``local'' or extragalactic objects) and \hii\ galaxies 
showing nebular \heii\ are considered to be quite rare (e.g.\ Garnett
\etal\ 1991). Although not intended as a complete inventory these 
authors quote only 17 objects drawn from the studies of dwarf emission-line 
galaxies by Kunth \& Sargent (1983) and Campbell \etal\ (1986).
Many new observations revealing nebular \heii\ emission have since
been found, mostly in the sample of Izotov and collaborators (see Sect.\
\ref{s_search}). 

In Table \ref{ta_heii} we list extragalactic \hii\ regions showing
\heii\ emission which is entirely attributed to nebular emission
processes. 
In some cases, however, WR signatures are suspected by some authors;
see column 8 for references.
We restrict ourself to objects from the Terlevich \etal\ (1991) 
catalogue analysed by various authors and the sample of Izotov 
and collaborators, since these constitute the largest available 
samples of such objects.
All except one object (Pox 105) from Kunth \& Sargent (1983) showing 
\heii\ emission was later classified as a WR galaxy (Kunth \& Joubert 1985
and references in Sect.\ \ref{s_catalogue}).
%
%ds Rajoute discussion objets Conti Table 2:
From the four emission line galaxies with narrow \heii\ emission listed by
Conti (1991, his Table 2) three are now classified as WR galaxies (see Table
\ref{ta_wrgals}); the presence of a broad \heii\ component due to WR stars
is suspected in the Seyfert 2 galaxy Pox 52 (Kunth \etal\ 1987).

From the 12 objects of Campbell \etal\ (1986) for which a measurement is 
%ds available, seven remain in this category, although three additional
% now Tol 1457-262B WR gal
available, six remain in this category, although two additional
objects are suspected to show broad \Heii.
In the other objects from Campbell \etal, WR features have been found by 
other investigators (see Sect.\ \ref{s_catalogue}), in part also in 
re-analysis of the same observational data (e.g.\ Masegosa \etal\ 1991).

%Campbell \etal\ (1986):
%objects with upper limit: not included.
%C 08-28A = Mrk 702: is WR gal
%T 1004-296= NGC 3125: more recent +better observations available
%T 1324-276=Tol 35: WR gal
%C 1543+091 = OVNI ??
%Mrk 36: WR gal accoring to IT98 

From the sample of Izotov and collaborators (including 60 objects published
up to 1998, i.e.\ including Izotov \& Thuan 1998), 40 show nebular \Heii.
However only 9 objects (listed in Table \ref{ta_heii}) reveal no broad 4686 
component according to these studies and the reanalysis of \git.

%Izotov sample: Total of 60 regions (from table. 1 missing compared to
%papers !!?)
%28 show WR, 9 show WN+WC.
%42 show nebular \Heii (18 show NO nebular \Heii),
%15 of those do NOT show WR signatures (i.e.\ listed in Table \ref{ta_heii} !).
%
%45 used for He abundance determination (others skipped for various 
%reasons...).
%1/4 of objects WITH nebular \Heii\ rejected;
%1/3 WITHOUT rejected.

%%%%%%%%%%%%%%%%%%%%%%%%%%%%%%%%%%%%%%%%%%%%%%%%%%%%%%%%%%%%%%%%%%%%%%%%
%\section{Unconventional objects with WR signatures}

%List Seyfert, LINERS, cD galaxies etc. (?) with
%possible or certain WR signatures are included in Table \ref{ta_catalogue}.

%Sey 2:
%Mrk 477, Mrk 1210, Mrk 309, NGC 5128

%LINER: NGC 6500, 6764, 3367

%%%%%%%%%%%%%%%%%%%%%%%%%%%%%%%%%%%%%%%%%%%%%%%%%%%%%%%%%%%%%%%%%%%%%%%%
%\section{Candidate WR galaxies}
\section{Suspected WR galaxies and galaxies without \heii}
\label{s_cand}

\subsection{Suspected WR galaxies}
Candidate WR galaxies resulting from some of the searches discussed
in Sect.\ \ref{s_search} or found loosely in the literature are
listed in Table \ref{ta_cand}.
The reader is referred to the original papers for the justification 
of the possible presence of WR stars.

Most of the suspected WR galaxies come from the spectrophotometric catalogue
of \hii\ galaxies of Terlevich \etal\ (1991), following the analysis
of Masegosa \etal\ (1991) or Pindao \etal\ (1999) which also includes new objects.
As stated earlier, from Masegosa \etal\ (1991) we take the subset 
of objects included in their Table 1 but not in Table 2 as candidates. 
Although partly based on the same observations the analysis of Masegosa \etal\ (1991)
and Pindao \etal (1999) do not always yield the same candidates.
We have retained all objects classified as ``candidate'' by either one
of these studies. 
New observations will be necessary to establish the definite presence
or absence of WR features.

Few objects from Vacca \& Conti (1992) have only an upper limit
on broad \heii. For most of them independent observations are now
available. Otherwise these objects are retained as suspected WR galaxies.

In some studies of IUE spectra strong UV P-Cygni lines of N and C 
(N~{\sc v} $\lambda$1240, C~{\sc iv} $\lambda$1550) have been interpreted as 
signatures of WR stars in theses objects (e.g. Durret \etal\ 1985, Lamb
\etal\ 1990).
However, these lines are also strong in O stars and hence cannot be used
as a clear diagnostic for WR stars (cf.\ Leitherer \etal\ 1995).
Therefore, objects suspected on these grounds have not been included in
Table \ref{ta_cand}.

In rare cases, WR stars have also been suspected on indirect grounds
(e.g.\  M100: Wozniak \etal\ 1998).

Interestingly, a broad WR bump is suggested to be present in the optical 
spectrum of two distant central cluster galaxies with strong cooling flows
(Abell 1068 and 1835, Allen 1995). The recent study of Contini \etal\
(1998), however, casts serious doubt about the reality of a broad
feature in Abell 1835.
If true, these objects with redshifts $z \sim$ 0.14 and 0.25 respectively,
represent the most distant objects known to date where WR stars have been 
detected from (rest-frame) optical spectra.
High redshift galaxies ($z \sim$ 3) may, however, also show WR signatures
(see Sect.\ \ref{s_conclude}).

%{\em IC 2574} ---
%Drissen \etal\ (1993) from imaging.

%{\em Mrk 496} ---
%Robledo-Rella \& Conti (1993) from spectra with low S/N  (S/N $<$ 10).

%{\em Case 184} ---
%Robledo-Rella \& Conti (1993) from low S/N spectra. 

%{\em Case 845} ---
%Robledo-Rella \& Conti (1993) from low S/N spectra. 

%{\em NGC 1068} ---
%Broad feature at $\sim$ 4660 \AA\ identified as  C\'{\sc iii} and/or 
%C\'{\sc iv} possibly due to WC in this Seyfert 2 galaxy (Evans \& Dopita
%1986).

%{\em M 100} ---
%ISO, Friedli, Wozniak ...

\subsection{Emission line galaxies without HeII}
Conti (1991, his Table 3) lists a sample of emission line galaxies which 
have properties similar to WR galaxies, and where a search for \heii\ 
emission (broad or narrow) has been made with negative results.
For obvious reasons such a list is necessarily incomplete and 
the inclusion in such a list also strongly depends
on the sensitivity (S/N, resolution etc.) of the data.
We therefore renounce on such a compilation. However, 
few updates are appropriate on some objects from Table 3 of Conti (1991).

{\em Mrk 1087       } ---
While \vc92\ provide only an upper limit on \heii, a detection is provided
in the recent spectra of Vaceli \etal\ (1997). No information on
the width of this line is given.
Retained as suspected WR galaxy.

{\em Mrk 1094       } ---
We have retained the criteria of \kj85\ and hence included this object
in Table \ref{ta_wrgals}. No new observations published.

{\em 0833+652 = IRAS 08339+6517} ---
From their spectrum Veilleux \etal\ conclude that no WR features are
present in this galaxy (=0833+652 in Conti 1991).

{\em Tol 2 = Tol 0957-278} ---
Same comment as for Mrk 1094.

{\em Tol 9 = Tol 1032-283}
Although no broad \heii\ feature seems present in this object (Kunth
\& Schild 1986) we list in the category of suspected WR galaxies
based on the possible detection of other broad features shortward
of 5876 \AA\ (Kunth \& Schild 1986).

Other objects for which an upper limit on broad WR features is given
or where a non-detection is signaled are found in publications
issued from the systematic searches discussed in Sect.\ \ref{s_search}.

%%%%%%%%%%%%%%%%%%%%%%%%%%%%%%%%%%%%%%%%%%%%%%%%%%%%%%%%%%%%%%%%%%%%%%%%
\section{Discussion and conclusion}
\label{s_conclude}
We have presented an up-to-date and presumably quite complete compilation of 
WR galaxies from the literature. The number of such objects has considerably
%dsnumber
increased in the last years and now totals 139 (Table \ref{ta_wrgals}).
In addition to {\em broad} \heii, the basic ``classification line'' for 
WR galaxies, we include, for the first time, relevant information about the 
presence of various other broad emission lines. In particular the presence of
\civ\ emission originating essentially from WC stars is now detected 
in many objects.

A large fraction of WR galaxies also show {\em nebular} \heii\ emission,
indicative of high excitation. 
Where available we also include information about this line. 
Conversely, few objects are known which show  {\em only nebular} \Heii\ emission,
i.e.\ no apparent signs of broad stellar emission features.
A list of these extra-galactic \hii\ regions is provided in Table \ref{ta_heii}.
This may in particular be used to investigate the possible link between
the phenomenon of high excitation and the presence of WR stars
as suggested e.g.\ by Schaerer (1996).

We have also compiled a list of objects suspected to harbour WR stars
(Table \ref{ta_cand}). This could serve for future follow-up spectroscopy.

Most of the work on WR galaxies and related objects discussed above
is based on spectroscopy in the visible.
{\em Are populations of WR stars also detectable at other wavelengths ?}

In the infrared no direct signature of WR populations have been detected
to the best of our knowledge.
The most likely explanations are that
{\em 1)} many strong features of WR stars coincide with strong nebular
        lines (e.g.\ He~{\sc i} 2.06$\mu$m, Br$\gamma$), and
{\em 2)} WR features in the IR are strongly diluted by cool stars, which contribute
        the bulk of the emission at these wavelengths.

The strongest indicator of WR stars in UV is the presence of broad \Heii\ 
$\lambda$1640 emission which is not seen in emission in other stars.
Predictions of its strength and the expected line profile for integrated 
populations are given by Leitherer \etal\ (1995) and Schaerer \& Vacca (1998).
The use of this line is, however, not straightforward for several reasons.
Potential difficulties specific to IUE spectra have been discussed by 
Leitherer \etal\ (1995).
\Heii\ $\lambda$1640 is detected in the average spectra of $\sim$ 20 starbursts 
(including several WR galaxies) of low and high metallicity of Heckman \etal\ (1998).
In particular the spectra illustrate also the presence of multiple stellar 
and interstellar absorption lines in this wavelength range, which
complicate the quantitative use of this line for diagnostics.

UV high resolution and high S/N spectra obtained with HST indeed show the 
presence of this emission line in known WR galaxies (e.g.\ NGC 4214, NGC 1741;
Conti \etal\ 1996, Leitherer \etal\ 1996).
Interestingly these objects show a close resemblance with recently discovered 
high redshift galaxies (cf.\ Steidel \etal\ 1996, Ebbels \etal\ 1996, Lowenthal 
\etal\ 1997). 
In addition to the strongest stellar UV lines (Si~{\sc iv}, C~{\sc iv} wind lines)
the average spectrum of Lowenthal \etal\ of 11 objects with $z \sim$ 3 also
shows the \Heii\ $\lambda$1640 emission line !
Quantitative studies of stellar populations including possibly WR stars should be
possible in the future using these lines.

The numerous recent findings of massive stars in Seyfert 2 and LINERs,
as well as detections of high redshift galaxies exhibiting signatures
of massive stars further stress the need to deepen our understanding of
the physical processes in ``local objects'', and illustrate the interest 
of studies on massive stars and their interactions with the ISM, 
stellar populations, and starbursts in a wider context.
It is the hope that our compilation will  provide a useful basis for 
such undertakings.

%%%%%%%%%%%%%%%%%%%%%%%%%%%%%%%%%%%%%%%%%%%%%%%%%%%%%%%%%%%%%%%%%%%%%%%%
\begin{acknowledgements}
We thank Rosa Gonz\'alez-Delgado, Daniel Kunth, Cristal Martin, and 
Bill Vacca for useful communications and discussions, and
Peter Conti, Daniel Kunth, Claus Leitherer and Yuri Izotov for
comments on an earlier version of the manuscript.
Yuri Izotov, Benoit Joguet, Bill Vacca, and Donna Weistrop, kindly provided results
before publication.
Some comments and corrections were made during the IAU Symposium 193 by
several colleagues. We thank them here for their help.
This research has made extensive use of the NASA Extragalactic Data\-base (NED),
the NASA Astrophysics Data System (ADS) Article Service, and SIMBAD,
which is operated by the CDS in Strasbourg, France.
DS acknowledges a grant from the Swiss National Foundation of Scientific
Research.

\end{acknowledgements}

%%%%%%%%%%%%%%%%%%%%%%%%%%%%%%%%%%%%%%%%%%%%%%%%%%%%%%%%%%%%%%%%%%%%%%%%

%%%%%%%%%%%%%%%%%%%%%%%%%%%%%%%%%%%%%%%%%%%%%%%%%%%%%%%%%%%%%%%%%%%%%%%%
\clearpage
\begin{table*}
\caption{List of WR galaxies (all WR populations outside Local Group).
(C91) stands for objects included in Conti (1991) catalogue.
\label{ta_wrgals}
}
%\begin{flushleft}
%{\scriptsize
{
\begin{tabular}{llllrrllllll}
%\begin{longtable}{llllrrllllll}
%\caption{\\ Table 1: List of WR galaxies (all WR populations outside Local Group).
%(C91) stands for objects included in Conti (1991) catalogue.} \\
\hline
\hline
        & R.A.   & Decl.      & Morphology & $m_{\rm b}$ & $V_{\rm rad}$ & Broad \heii\ & Nebular & \civ\       \\ % & Comment      \\
Name    & [1950] &   [1950]   &            & [mag]       & [\kms]        & reference    & \heii\  &reference            \\
%       & R.A.   & Decl.      & Morphology & $m_{\rm b}$ & $V_{\rm rad}$ & Broad  & Neb.    & 5808       & Comment      \\
%Name    & [1950] &   [1950]   &            & [mag]       & [\kms]        & \heii\ & \Heii\  &          \\
\hline
\noalign{\smallskip}                          
NGC 53         & 00 12 16.1     & -60 36 19   & (R')SB(r)ab & 13.3 & 4568 & MMO91      &   &                 \\ %& MAX ?  
NGC 55         & 00 12 38.0     & -39 29 54   & SB(s)m: sp  &  8.4 &  129 & RD86       &   & RD86       \\    
UM 228         & 00 18 27.2     & +00 36 09   &             & 17.0 &29490 & MMO91,P98           \\
UM 48          & 00 33 35.8     & +04 21 37   & S           & 15.5 & 4907 & MMO91      &   &            \\ % (=UGC 359)
Mrk 960        & 00 46 04.8     & -12 59 21   & (R)SB0 pec? & 13.5 & 6407 & KC98       &-- &            \\
NGC 300        & 00 52 31.7     & -37 57 15   & SA(s)d      &  8.7 &  144 & DRW83      &   & DRW83      \\    
IRAS 01003-2238& 01 00 23.6     & -22 38 03   &             &  18.9&35310 & AHM88,(C91)&   &            \\ 
UM 311         & 01 13 00.5     & -01 07 22   &             &      & 1798 & MMO91      & ! & IT 98      \\
Tol 0121-376   & 01 21 55.8     & -37 37 55   &             &      &10500 & MMO91,P98           \\
Mink           & 01 23 14.2     & -01 37 54   & Irr         & 17.0 & 5638 & B85,(C91)  &UL &            \\ %data from LEDA 
Mrk 996        & 01 25 04.5     & -06 35 08   &             & 15.5 & 1622 & TIL96      &-- & TIL96      \\
Mrk 589        & 02 11 08.7     & +03 52 08   & S?          & 14.5 & 3436 & GIT98      &-- & GIT98      \\
UM 420         & 02 18 20.4     & +00 19 42   & Compact     & 16.5 &17652 & IT98       & ! & GIT98      \\ 
Mrk 1039       & 02 25 07.2     & -10 23 19   & Sc? sp; HII & 15.5 & 2089 & H98        & Q & H98        \\ 
Tol 0226-390   & 02 26 10.0     & -39 02 39   &             & 15.3 &14340 & MMO91,P98               \\
Tol 0242-387   & 02 42 39.2     & -38 47 31   & HII         & 17.8 &37807 & MMO91,P98           \\ 
Mrk 598        & 02 43 52.2     & +07 11 34   &             & 17.0 & 5396 & P98                 \\
NGC 1140       & 02 52 08.0     & -10 13 49   & IBm pec     & 12.8 & 1501 & GIT98      &-- & GIT98              \\
NGC 1156       & 02 56 46.8     & +25 02 21   & IB(s)m      & 12.3 &  375 & HFW95      &   & \\  
NGC 1313       & 03 17 39.0     & -66 40 42   & SB(s)d      &  9.2 &  475 & WR97       &   & WR97       \\ 
NGC 1365       & 03 31 41.8     & -36 18 27   & (R')SBb(s)b & 10.3 & 1636 &            &   & PC92       \\ 
SBS 0335-052   & 03 35 15.1     & -05 12 26   &             & 17.0 & 4043 & I98        & ! & \\
NGC 1510       & 04 01 53.9     & -43 32 14   & SA0 pec?    & 13.5 &  913 & EN84,(C91) &   & \\ 
NGC 1569       & 04 26 04.6     & +64 44 23   & IBm         & 11.9 &$-$104& DR94       &   & \\ 
NGC 1614       & 04 31 35.5     & -08 40 56   & SB(s)c pec  & 13.6 & 4778 & P98,(C91)          \\
VII Zw 19      & 04 35 41.3     & +67 38 19   &             & 16.0 & 4830 & KJ85,(C91) &   & \\
NGC 1741 (B)   & 04 59 06.5     & -04 20 08   & SB(s)m pec  & 15.2 & 4171 & KS86,(C91) & ? & &  \\ %part of H31 compact group
H31A           & 04 59 09.9     & -04 19 52   & Sdm         & 15.6 & 4042 & R90,(C91)  &   & \\
Mrk 1094       & 05 08 17.4     & -02 44 33   & I0 pec?     & 14.1 & 2831 & KJ85       &   & \\
II Zw 40       & 05 53 04.9     & +03 23 07   & BCD;Irr     & 15.5 &  789 & KS81,(C91) & ! & GIT98      \\
Tol 0633-415   & 06 33 35.0     & -41 31 12   &             & 16.5 & 5096 & MMO91      &   & \\
Mrk 5          & 06 35 24.4     & +75 40 22   & I?          & 15.6 & 792  & IT98       &   & \\
IRAS 07164+5301& 07 16 28.6     & +53 01 06   &             &      &      & H96        &   & \\    
Mrk 1199       & 07 20 28.3     & +33 32 21   & Sc          & 13.7 & 4107 & IT98       & Q & IT98       \\
NGC 2363       & 07 23 23.7     & +69 17 33   & Sm; HII     & 11.6 & 107 & G94         & Q & G94  \\ %HII region in NGC 2366
Mrk 8          & 07 23 36.8     & +72 13 58   & S?          & 13.8 & 3496 & KJ85,(C91) &   & \\
NGC 2403       & 07 32 05.5     & +65 42 40   & SAB(s)cd    & 8.93 & 131  & DR96       &   & DR96 \\ 
VII Zw 187     & 07 49 27.4     & +72 24 57   &             & 17.4 &      & KJ85       &   & \\ 
SBS 0749+582   & 07 49 48.0     & +58 16 00   &             & 19.0 & 9548 & ITL97      &-- & \\
Mrk 1210       & 08 01 27.0     & +05 15 22   & S?          & 14.3 & 4046 & SCS98      & ! &           \\ %& Sey 2  
IRAS 08208+2816& 08 20 52.2     & +28 15 57   & Irr         & 15.5 &14034 & H98        &   & H98        \\
He 2-10        & 08 34 07.0     & -26 14 06   & I0? pec     & 12.4 &  873 & AWG76,(C91)&   & VC92\\
Mrk 702        & 08 42 45.1     & +16 16 44   & Compact     & 15.7 &15840 & MMO91      &-- & GIT98  \\ 
SBS 0907+543   & 09 07 31.8     & +54 23 08   &             & 17.0 & 8124 & ITL97      & Q & \\
SBS 0926+606   & 09 26 20.0     & +60 40 02   & BCG         & 17.5 & 4122 & ITL97      & Q & \\
I Zw 18        & 09 30 30.2     & +55 27 49   & Compact     & 15.6 &  742 & I97        & Q & I97,L97    \\
ESO 566-7      & 09 42 38.0     & -19 29 42   & SBb pec?    & 15.3 & 9890 & MMO91      &   &            \\ 
NGC 3003       & 09 45 38.5     & +33 39 19   & Sbc?        & 12.3 & 1478 & HFW95      &   & \\
Mrk 22         & 09 46 03.3     & +55 48 48   &             & 15.7 & 1499 & ITL94      & Q & GIT98      \\
Mrk 1236       & 09 47 19.9     & +00 51 01   & SABcd       & 13.5 & 1829 & KS86,(C91) & ! & GIT98      \\
SBS 0948+532   & 09 48 10.2     & +53 13 41   &             & 18.0 &13890 & ITL94      & Q & GIT98  \\ %& Sey in NED
\noalign{\smallskip}                          
\hline                                        
\hline                                        
\end{tabular}
}
\end{table*}
\clearpage

\begin{table*}
%\caption{{\bf Table 1:} List of WR galaxies (cont.)}
\centerline{\small {\bf Table 1:} List of WR galaxies (cont.)}
{
\begin{tabular}{llllrrllllll}
\hline
\hline
        & R.A.   & Decl.      & Morphology & $m_{\rm b}$ & $V_{\rm rad}$ & Broad \heii\ & Nebular & \civ\       \\ % & Comment      \\
Name    & [1950] &   [1950]   &            & [mag]       & [\kms]        & reference    & \heii\  &reference            \\
\hline
\noalign{\smallskip}            
NGC 3049       & 09 52 10.2     & +09 30 32   & SB(rs)ab    & 13.0 & 1494 & KS86,(C91) &-- & SCK99      \\
Mrk 712        & 09 53 59.1     & +15 52 34   & SB          & 14.5 & 4560 & CDS95      & ? &  \\
Tol 0957-278   & 09 57 05.0     & -27 53 30   & Multiple?   & 14.4 &  710 & KJ85       &   & \\
NGC 3125       & 10 04 18.2     & -29 41 29   & S           & 13.5 &  865 & KS81,(C91) & ?  & SCK99     \\
Tol 1025-284   & 10 25 12.0     & -28 26 00   &             & 16.9 & 9593 & P98                 \\
Mrk 33         & 10 29 22.7     & +54 39 31   & Im pec      & 13.4 & 1461 & KJ85,(C91) &   & \\
Mrk 1434       & 10 30 56.3     & +58 19 20   & BCG         & 16.5 & 2269 & ITL97      & Q & GIT98      \\
Mrk 1259       & 10 36 03.0     & -06 54 37   & S0 pec?     & 13.5 & 2159 & OTT97      &-- & GIT98      \\
Mrk 724        & 10 38 26.8     & +21 37 24   & Compact     & 16.5 & 1139 & KS86,(C91) & ! & KS86       \\
NGC 3353       & 10 42 15.8     & +56 13 26   & BCD/Irr     & 13.2 &  944 & S96,H98    & Q & S96,H98    \\
NGC 3367       & 10 43 55.4     & +14 00 58   & SB(rs)c     & 12.1 & 3037 & HFW95      &   &   \\ %& LINER in NED 
NGC 3395       & 10 47 02.7     & +33 14 44   & SAB(rs)cd   & 12.4 & 1625 & W98        &-- & W98        \\
NGC 3396       & 10 47 08.9     & +33 15 18   & IBm pec     & 12.6 & 1625 & W98        & ! & W98        \\
Mrk 1271       & 10 53 33.3     & +06 26 24   & Compact     & 14.8 & 1049 & IT98       & Q &  ?GIT98 \\
SBS 1054+365   & 10 54 59.8     & +36 31 30   &             & 16.0 &  603 & ITL97      &-- &-- \\
Mrk 36         & 11 02 15.6     & +29 24 34   & BCD         & 15.5 &  646 & IT98       & Q & \\
%ds rajoute Arp 299B+C d'une private communication... --> part of NGC 3690 -- mention in text only!
%Arp 299C       & 11 25 37.8     & +58 51 14   & AGN         & 16   & 3111 & V96        &   &           \\
%Arp 299B       & 11 25 41.5     & +58 50 13   & SBm? pec    & 11.8 & 3123 & V96               &   &            \\
NGC 3690       & 11 25 42.4     & +58 50 17   & IBm pec     & 12.0 & 3121 & HFW95      &   & \\
%wrong coord in C91 !Mrk 178        & 10 30 45.9     & +49 30 52   & pair        & 14.4 &  249 & GRZ88,(C91)& ! & GIT98      \\
Mrk 178        & 11 30 45.9     & +49 30 52   & pair        &      &  249 &  GRZ88,(C91)& ! & GIT98      \\
NGC 3738       & 11 33 04.2     & +54 48 04   & Im          & 12.1 &  229 & M97        &   & \\
UM 439         & 11 34 03.0     & +01 05 36   &             & 15.1 & 1169 & MMO91      &   & \\
Mrk 182        & 11 34 17.7     & +20 12 12   & Compact     & 17.0 & 6328 & GIT98      & ? & -- \\
Mrk 1450       & 11 35 51.3     & +58 09 04   & Compact     & 15.5 &  946 & ITL94      & Q & ITL94      \\
Mrk 1304       & 11 39 38.6     & +00 36 42   & Sb pec      & 14.7 & 5488 & MMO91      & ! & GIT98      \\
Mrk 1305       & 11 40 24.6     & -08 03 18   &             & 15.5 & 2997 & GIT98      &-- & GIT98      \\
Mrk 750        & 11 47 28.0     & +15 18 04   & BCD         & 15.8 &  754 & KJ85,(C91) & Q & GIT98      \\
%Tol 1148-2020  & 11 48          & -20 20.0    & MMO91      &   & & = Pox 4 = C 1148-203 ??? cf. Campell et al.86 \\
Pox 4          & 11 48 39.0     & -20 19 17   &             & 16.2 & 3589 & KJ85       & ? & &          \\
UM 461         & 11 48 59.4     & -02 05 41   & BCD/Irr     & 16.9 &  899 & C91        &   & \\
Mrk 1307       & 11 50 03.8     & -02 11 28   & Pec;BCD     & 14.5 & 1012 & IT98       & Q & \\
Mrk 193        & 11 52 52.1     & +57 56 26   & Compact     & 16.5 & 5282 & GIT98      & ! & \\
ISZ 59         & 11 54 54.7     & -19 20 44   & S0          & 16.8 & 1781 & KJ85,(C91) &   & \\
NGC 3995       & 11 55 10.4     & +32 34 24   & SAm pec     & 12.7 & 3254 & W98        & ! & W98        \\
NGC 4038       & 11 59 19.0     & -18 35 12   & SB(s)m pec  & 10.9 & 1642 & RD86       &   & \\    
SBS 1211+540   & 12 11 33.9     & +54 01 58   &             & 17.8 &  929 & GIT98      & ! & \\
NGC 4214       & 12 13 08.3     & +36 36 22   & IAB(s)m     & 10.2 &  291 & SF91       &   & SF91, MHK91 \\
NGC 4216       & 12 13 21.5     & +13 25 40   & SAB(s)b     & 11.0 &  131 & RD86       &   & \\    %& HII/LINER in NED 
NGC 4236       & 12 14 21.8     & +69 44 36   & SB(s)dm     & 10.1 &    0 & GP94       &   & \\
M 106          & 12 16 29.4     & +47 34 53   & SAB(s)bc    & 9.10 &  448 & C98          &   &     \\ %& Sy1.9 in NED 
SBS 1222+614   & 12 22 44.5     & +61 25 46   &             & 17.0 &  734 & ITL97      & Q & ITL97      \\
NGC 4385       & 12 23 09.0     & +00 50 57   & SB(rs)0+    & 13.2 & 2140 & CS86,(C91) &   & CS86  \\ %& MAX: cat Terlevich
II Zw 62       & 12 23 38.4     & +07 56 37   &             & 17.2 & 3930 & KJ85,(C91) &   & \\
Mrk 209        & 12 23 50.5     & +48 46 13   & Sm pec      & 15.2 &  281 & ITL97      & Q & GIT98      \\
NGC 4449       & 12 25 45.9     & +44 22 16   & IBm         & 10.0 &  207 & MK97       &   & MK97       \\
NGC 4532       & 12 31 46.7     & +06 44 39   & IBm         & 12.3 & 2012 & HFW95      &   & \\
Mrk 1329       & 12 34 29.8     & +07 12 01   & SBmIII      & 14.4 & 1632 & GIT98      &-- & ?GIT98     \\
Tol 1235-350   & 12 35 46.0     & -35 02 42   & S0:         & 17.1 & 2998 & P98                 \\
NGC 4670       & 12 42 49.8     & +27 23 54   & SB(s)0/a pec& 13.1 & 1069 & MHK91      &   & \\
Tol 1247-232   & 12 47 39.0     & -23 17 38   &             & 15.5 &14400 & MMO91      &   & \\
SBS 1249+493   & 12 49 35.8     & +49 19 45   & BCG         & 17.5 & 7330 & GIT98      & ! & \\
NGC 4861       & 12 56 39.7     & +35 07 50   & SB(s)m      & 12.9 &  847 & DS86,(C91) & Q & DS86       \\
Tol 30         & 13 03 03.0     & -28 09 12   & SB          & 15.5 &      & C96        & ? & \\
Pox 120        & 13 04 04.7     & -11 48 20   &             & 15.7 & 6220 & KJ85,(C91) &   & \\
\noalign{\smallskip}                          
\hline                                        
\hline                                        
\end{tabular}
}
\end{table*}
\clearpage

\begin{table*}
%\caption[]{{\bf Table 1:} List of WR galaxies (cont.)}
\centerline{\small {\bf Table 1:} List of WR galaxies (cont.)}{
\begin{tabular}{llllrrllllll}
\hline
\hline
        & R.A.   & Decl.      & Morphology & $m_{\rm b}$ & $V_{\rm rad}$ & Broad \heii\ & Nebular & \civ\       \\ % & Comment      \\
Name    & [1950] &   [1950]   &            & [mag]       & [\kms]        & reference    & \heii\  &reference            \\
\hline
\noalign{\smallskip}            
Pox 139        & 13 09 20.8     & -11 47 54   & SB(s)d      & 15.0 & 2107 & KJ85,(C91) &   & \\
NGC 5068       & 13 16 13.0     & -20 46 36   & SB(s)d      & 10.7 &  673 & RD86       &   & \\    
SBS 1319+579   & 13 19 25.2     & +57 57 09   &             & 18.5 & 2060 & ITL97      & Q & GIT98      \\
NGC 5128       & 13 22 31.6     & -42 45 33   & S0 pec      &  7.8 &  547 & M81,RD86   &   & M81 \\ % & Sey2 in NED     
Pox 186        & 13 23 12.0     & -11 22 00   &             & 17.0 & 1170 & KJ85,(C91) &   & \\
Tol 35         & 13 24 20.0     & -27 41 48   & S?          & 14.4 & 1814 & KJ85,(C91) & ? & \\
M 83           & 13 34 11.5     & -29 36 42   & SAB(s)c     &  8.2 &  516 & RD86       &   & \\    
NGC 5253       & 13 37 05.1     & -31 23 13   & IM pec      & 10.9 &  404 & CTM86,(C91)& ? & SCK99      \\
Mrk 67         & 13 39 39.6     & +30 46 16   & BCD/Irr     & 16.5 &  958 & C91        &   & \\
Mrk 1486       & 13 58 09.4     & +57 40 54   &             &      &10143 & ITL97      & Q & \\
Tol 89         & 13 58 26.0     & -32 49 20   & SB(rs)dm    & 16.0 & 1216 & DBB85,(C91)& ? & SCK99 \\ %& HII region in NGC 5398
NGC 5430       & 13 59 08.5     & +59 34 16   & SB(s)b      & 12.7 & 3028 & K82,(C91)  &   & \\
NGC 5408       & 14 00 17.5     & -41 08 19   & IB(s)m      & 12.2 & 509  & MMO91      & ! &  \\ %& MAX ??
NGC 5457       & 14 01 26.3     & +54 35 18   & SAB(rs)cd   &  8.3 &  241 & R82,DRW83  &   & \\    
NGC 5461       & 14 01 54.9     & +54 33 24   &             &      &  298 & R82        &   &  \\ %& GHII in M 101   
NGC 5471       & 14 02 43.4     & +54 38 08   &             & 15.0 &  297 & MHK91      & ? &  \\ %& GHII in M 101
SBS 1408+551A  & 14 08 14.0     & +55 11 00   &             & 18.0 &23190 & I96        & Q & I96        \\
CGCG 219-066   & 14 15 03.6     & +43 44 04   &             & 15.6 &  649 & I98        & ! & \\
Mrk 475        & 14 37 03.6     & +37 01 13   & BCD         & 14.5 &  540 & C91        & Q & ITL94      \\
Mrk 477        & 14 39 02.5     & +53 43 04   &             &      &11340 & H97        & ! &        \\ %& Sey2 
Tol 1457-262A  & 14 57 31.8     & -26 14 40   & HII         &      & 5180 & C96        & - &  \\ 
Tol 1457-262B  & 14 57 32.1     & -26 14 49   & HII         & 14.7 & 5249 & P98        & ? &  \\
SBS 1533+574B  & 15 33 04.1     & +57 27 00   &             &      & 3390 & GIT98      & ! & GIT98      \\
IC 4662        & 17 42 12.0     & -64 37 18   & IBm         & 11.7 &  308 & RD86       &   & RR91       \\
NGC 6500       & 17 53 48.1     & +18 20 41   & SAab        & 13.1 & 3003 & B97        &   &   \\ %& LINER 
Fairall 44     & 18 09 19.3     & -57 44 48   & S? pec      & 14.8 & 4948 & KC98       & Q & KC98 &     \\
NGC 6764       & 19 07 01.2     & +50 51 08   & SB(s)bc     & 12.6 & 2416 & OC82,(C91) &   &     \\ %Sey2/LINER 
Tol 1924-416   & 19 24 29.0     & -41 40 36   & pec HII     & 13.3 & 2874 & P98        & ? &     \\
IC 4870        & 19 32 48.0     & -65 55 30   & IBm? pec Sy2& 13.9 & 889  & JK99       & ! & \\
IC 5154        & 22 00 41.7     & -66 21 25   & Irr Sy2     & 14.9 & 3118 & JK99       & ? & JK99 \\
ESO 108-IG 017 & 22 07 00.3     & -67 06 59   & I0? Sy2     & 14.4 & 2191 & JK99       & ? & JK99 \\
Mrk 309        & 22 50 10.0     & +24 27 52   & Sa          & 15.4 &12645 & OC82,(C91) &   & OC82 \\  %& Sey2 in NED
Mrk 315        & 23 01 35.7     & +22 21 16   & E1 pec?     & 14.8 &11661 & KJ85       &   & \\
ESO 148-IG 02  & 23 12 51.0     & -59 19 40   & Merger      & 15.2 &13380 & JB88,(C91) &   & \\
III Zw 107     & 23 27 40.4     & +25 15 27   & Sb          & 15.6 & 6176 & KJ85,(C91) &   & \\
Mrk 930        & 23 29 29.2     & +28 40 16   &             &      & 5400 & IT98       & ! & GIT98      \\
NGC 7714       & 23 33 40.6     & +01 52 42   & SB(s)b pec  & 13.0 & 2798 & B85,(C91)  & Q & G97        \\
\noalign{\smallskip}                          
\hline                                        
\hline                                        
\noalign{\smallskip}                          
\noalign{\small 
KEY TO column 8 (nebular \Heii). ---
no entry:       no information available, 
``--'':         absent,
``?'':          possible/suspected, 
``!'':          present, but no data available,
``UL'':         upper limit available,
``Q'':          measurement available 
}
\noalign{\small 
KEY TO REFERENCES. ---
AHM88   Armus et al. 1988,              
AWG76   Allen et al. 1976,              
B85     Breugel et al. 1985,               
B97     Barth \etal\ 1997,              
C91     Conti 1991,                      
C96     Contini 1996,                   
C98     Castellanos et al.\ 1998,
CDS95   Contini et al. 1995,             
CS86    Campbell \& Smith 1986,         
CTM86   Campbell et al. 1986,           
DBB85   Durret et al. 1985,             
DR94    Drissen \& Roy 1994,             
DR96    Drissen \& Roy 1996,             
DRM93   Drissen et al. 1993,             
DRW83   D'Odorico et al. 1983,          
DS86    Dinerstein \& shields 1986,           
EN84    Eichendorf \& Nieto 1984,        
G97     Garc\'{\i}a-Vargas \etal\ 1997, 
GIT98   Guseva \etal\ 1998,             
GP94    Gonzalez-Delgado \& Perez 1994,     
GRZ88   Gonzalez-Riestra et al. 1988,   
H96     Huang et al. 1996,               
H97     Heckman \etal\ 1997,            
H98     Huang et al. 1998,               
HFW95   Ho et al. 1995,                  
I96     Izotov et al. 1996,              
I97     Izotov et al. 1997b,              
I98     Izotov et al. 1998,              
IT98    Izotov \& Thuan 1998,            
ITL94   Izotov et al. 1994,              
ITL97   Izotov et al. 1997a,              
JB88    Johansson \& Bergvall 1988,
JK99    Joguet \& Kunth 1999,
K82     Keel 1982,
KC98    Kovo \& Contini 1998,
KJ85    Kunth \& Joubert 1985,           
KS81    Kunth \& Sargent 1981,            
KS86    Kunth \& Schild 1986,            
%KMH94   Kunth \& Mas-Hesse 1994,        
L97     Legrand et al. 1997,             
%LBT90   Lamb et al. 1990,                
M81     Moellenhoff 1981,                
M97     Martin 1997,                     
MK97    Martin \& Kennicut 1997,         
MMO91   Masegosa et al. 1991,            
MHK91   Mas-Hesse \& Kunth 1991,         
OC82    Osterbrock \& Cohen 1982,
OTT97   Ohyama et al. 1997,              
P98     Pindao 1998,
PC92    Phillips \& Conti 1992,           
R82     Rayo et al. 1982,                
RR91    Richter \& Rosa 1991,
R90     Rubin et al. 1990,
RD86    Rosa \& D'Odorico 1986,          
S97     Schaerer et al. 1997,
S96     Steel et al. 1996,               
SCK99   Schaerer \etal\ 1999,
SCS98   Storchi-Bergmann \etal\ 1998,
SF91    Sargent \& Fillipenko 1991,      
TIL96   Thuan et al. 1996,               
V88     Vilchez et al. 1988,             
V96     Vacca 1996,
VC92    Vacca \& Conti 1992,             
W98     Weistrop \etal\ 1998,
WR97    Walsh \& Roy 1997               
}
\end{tabular}
}
\end{table*}

%%%%%%%%%%%%%%%%%%%%%%%%%%%%%%%%%%%%%%%%%%%%%%%%%%%%%%%%%%%%%%%%%%%%%%%%
\clearpage
\begin{table*}
\caption[]{List of extragalactic \hii\ regions with nebular \Heii.}
%\begin{flushleft}
%{\scriptsize
{
\begin{tabular}{llllrrllllll}
\hline
\hline
        & R.A.   & Decl.      & Morphology & $m_{\rm b}$ & $V_{\rm rad}$ & Nebular \heii\ & Comment   \\ 
Name    & [1950] &   [1950]   &            & [mag]       & [\kms]        & reference      &             \\
\hline
\noalign{\smallskip}                       
%
% NOTE: coordonees deja prised de NED
%   
SBS 0749+568   & 07 49 37.7     & +56 49 48   & BCG         & 18.0 & 5471 & ITL97       &       \\
Mrk 1416       & 09 17 25.9     & +52 46 50   & Irr         & 17.0 & 2305 & ITL97       &       \\
Tol 4          & 10 08 31.0     & -28 39 18   & E?          & 14.4 & 4219 & CTM86       & also C96; P99: WR?    \\ %Tol4=Tol 1008-286 
SBS 1116+583B  & 11 16 31.4     & +58 20 16   &             & 19.5 & 9905 & ITL97       &       \\ % noisy spectrum in red
UM 469         & 11 54 38.6     & +02 45 10   &             & 18.0 &17388 & C96         &       \\
SBS 1159+545   & 11 59 29.0     & +54 32 33   & BCG         & 18.0 & 3537 & ITL94       &       \\ % noisy red ?        
SBS 1205+557   & 12 05 57.5     & +55 42 07   & BCG         & 15.5 & 1751 & ITL97       &       \\
Tol 21         & 12 14 42.0     & -27 45 00   &             & 17.5 & 7795 & CTM86       &       \\ % = T 1214-277
Mrk 1318       & 12 16 36.5     & +04 07 58   & E pec       & 14.0 & 1526 & C96         & P99: WR?      \\
Fairall 30     & 12 36 32.0     & -39 54 42   & SAB(r)0     &      & 1199 & CTM86       &       \\
UM 533         & 12 57 25.0     & +02 19 08   & dIn         & 15.7 &  874 & KC98        &       \\ %dsnew
Pox 105        & 13 00 00.0     & -11 10 00   &             & 17.0 & 3405 & KS83,KJ85   &       \\
Tol 1304-386   & 13 04 32.0     & -38 38 48   &             &      & 4197 & CTM86       & P99: WR?      \\
Tol 78         & 13 04 48.0     & -35 22 00   &             & 17.0 & 4197 & CTM86       &       \\ % = T 1304-353
Tol 111        & 13 45 24.0     & -42 06 00   &             & 16.3 & 2398 & CTM86       &       \\
SBS 1420+544   & 14 20 59.1     & +54 27 42   & BCG         & 18.0 & 6176 & TIL95       &       \\
CG 1258        & 14 41 50.9     & +29 28 32   &             & 18.0 &13614 & ITL97       &       \\
SBS 1533+469   & 15 33 24.0     & +46 59 00   &             &      & 5666 & TIL95       &       \\
HS 1851+6933   & 18 51 38.6     & +69 33 16   &             & 17.0 & 7495 & I96         &       \\
\noalign{\smallskip}                          
\hline                                        
\hline                                        
\noalign{\smallskip}                          
\noalign{\small 
KEY TO REFERENCES. ---
C96     Contini 1996,
CTM86   Campbell et al. 1986, 
I96     Izotov \etal\ 1996,
I97c    Izotov \etal\ 1997c,
ITL94   Izotov \etal\ 1994,
ITL97   Izotov \etal\ 1997a,
KC98    Kovo \& Contini 1998,
KJ85    Kunth \& Joubert 1985,           
KS83    Kunth \& Sargent 1983,
P98     Pindao 1998,
P99     Pindao \etal\ 1999,
TIL     Thuan \etal\ 1995

}
\end{tabular}
}
%\end{flushleft}
\label{ta_heii}
\end{table*}
%%%%%%%%%%%%%%%%%%%%%%%%%%%%%%%%%%%%%%%%%%%%%%%%%%%%%%%%%%%%%%%%%%%%%%%%
%\clearpage
\begin{table*}
%\caption[]{List of candidate WR galaxies}
\caption[]{List of suspected WR galaxies}
%\begin{flushleft}
%{\scriptsize
{
\begin{tabular}{llllrrllllll}
\hline
\hline
        & R.A.   & Decl.      & Morphology & $m_{\rm b}$ & $V_{\rm rad}$ & Reference         \\ %& Comment  
Name    & [1950] &   [1950]   &            & [mag]       & [\kms]        &                      &               \\
\hline
%
% MMO91 objects: coords from NED (in general)
Fairall 4      & 00 12 49.2     & -57 31 22   & (R)SA(r)b   & 15.1 & 9796 & P99                 \\
Tol 0030-388   & 00 30 38.3     & -38 50 29   &             &      &13800 & MMO91               \\
Fairall 0007   & 00 42 06.0     & -60 07 00   &             &      &      & MMO91               \\
UM 080         & 00 57 46.8     & +04 23 46   &             & 17.8 & 4797 & P99                 \\ 
EQ 0102-310    & 01 02          & -31.0       &             &      &      & MMO91               \\ %=C0102-310 name after SIMBAD...
Tol 0127-397   & 01 27 02.4     & -39 46 03   &             &      & 4797 & P99                 \\ %dsnew
UM 354         & 01 37 11.1     & +01 10 57   &             &      & 9294 & P99                 \\
UM 377         & 01 52 18.8     & +01 02 28   &             &      & 8694 & P99                 \\
UM 396         & 02 04 51.0     & +02 42 41   &             &      & 6296 & P99                 \\
NGC 848        & 02 07 50.4     & -10 33 25   & (R')SB(s)ab & 13.6 & 4001 & GIT98       \\ %&-- & -- & SPECTRE pas clair !\\
NGC 1068       & 02 40 07.0     & -00 13 32   & (R)SA(rs)b  &  9.6 & 1136 & ED86                \\  %& Sey2 = M 77 
CAM 0357-3915  & 03 57 22.1     & -39 14 51   & HII         &      &22545 & P99                 \\
Tol 0440-381   & 04 40 23.3     & -38 06 41   &             &      &12300 & MMO91,P99           \\
Tol 0513-393   & 05 13 39.7     & -39 20 59   &             &      &15000 & P99                 \\
Tol 0538-416   & 05 38 32.9     & -41 38 37   & HII         &      & 13491& P99                 \\  %dsnew
Tol 0559-393   & 05 59 04.9     & -39 19 08   & Irr         & 15.0 &24915 & MMO91,P99           \\
Tol 0620-386   & 06 30 08.3     & -38 39 11   &             &      &20100 & P99         \\
Tol 0645-376   & 06 45 07.0     & -37 40 02   &             &      & 7795 & MMO91               \\
NPM1G+16.0158  & 08 41 01.4     & +16 10 38   &             & 18.3 &      & P99                 \\
ESO 566-8      & 09 42 39.2     & -19 28 55   & S pec       & 16.3 & 9761 & C96                 \\ 
Mrk 709        & 09 46 33.2     & +17 06 44   & BCD         & 17.0 & 1197 & P99                 \\
Tol 1021-289   & 10 21 22.0     & -28 58 30   & Interacting & 16.9 &17658 & P99                 \\ %dsnew
Tol 1025-285   & 10 25 06.2     & -28 32 13   & N           & 16.2 & 8694 & P99                 \\ %dsnew
Abell 1068     & 10 37 51.3     & +40 12 51   & cD          & 16.0 &41580 & A95                 \\ %& Central Cluster Galaxy
Tol 1032-283   & 10 32 18.4     & -28 19 52   & E4:         & 14.4 & 3190 & KS86                \\
Mrk 1301       & 11 33 10.9     & +35 36 43   & S0/a        & 14.5 & 1603 & RC93   \\ %& = Case 845  
UM 456         & 11 48 02.5     & -00 17 21   & Pec         & 15.9 & 1757 & P99                 \\ %dsnew
Pox 52         & 12 00 22.9     & -20 39 21   &             & 17.2 &      & KSB87               \\
UM 482         & 12 09 29.4     & -00 19 43   & Sm          & 16.2 &10544 & P99                 \\ %dsnew
UM 483         & 12 09 41.0     & +00 21 01   &             & 16.7 & 2099 & P99                 \\ %dsnew
M100           & 12 20 22.9     & +16 05 58   & SAB(s)bc    &  7.4 & 1571 & WO98                \\ %& LINER/HII 
NGC 4509       & 12 30 38.9     & +32 22 02   & Sab pec?    & 14.1 &  937 & RC93    \\ %& = Case 184
Tol 1258-363   & 12 58 19.0     & -36 19 54   &             &      & 4887 & P99                 \\
UM 594         & 13 35 32.7     & +00 16 26   & HII?        & 15.8 & 6653 & P99                 \\
NGC 5257 (W)   & 13 37 19.7     & +01 05 33   & SAB(s)b pec & 12.9 & 6798 & MMO91               \\ %=UM598 West
NGC 5291       & 13 44 33.3     & -30 09 31   & E pec:      & 15.1 & 4386 & DM98                \\ %dsnew
Tol 1345-419   & 13 45 14.2     & -41 55 23   & Sy2         & 17.0 & 11572 & P99                \\ %dsnew
UM 618         & 13 50 03.1     & +00 16 07   &             & 18.4 & 4197 & P99                 \\ %dsnew
Abell 1835     & 13 58 30.0     & +03 06 00   & cD          & 19.3 &75690 & A95                 \\ %& Central Cluster Galaxy 
EQ 1409-120    & 14 09          & +01 20      &             &      &      & MMO91               \\ %in SIMBAD, not NED
Cam 1409+1200  & 14 09 14.3     & +11 59 33   &             &      &16800 & P99         \\
NGC 6090       & 16 10 24.0     & +52 35 04   & Sd pec      & 14.5 & 8822 & RC93 \\ %& = Mrk 496
Tol 2006-393   & 20 06 39.7     & -39 22 32   &             &      & 9294 & P99                 \\
Tol 2138-405   & 21 38 14.8     & -40 32 35   &             &      &16800 & P99         \\
Tol 2146-391   & 21 46 45.0     & -39 08 10   &             &      & 8994 & MMO91,P99           \\
Tol 2259-398   & 22 59 34.0     & -39 49 42   & Compact     & 15.6 & 8858 & MMO91,P99           \\ %wrong ID =Tol 2259-3950
UM 159         & 23 19 31.9     & +01 12 24   & HII         &      & 8694 & P99                 \\ %dsnew
UM 160 W       & 23 21 46.0     & -00 23 30   &             &      & 2400 & P99                 \\ %dsnew
\noalign{\smallskip}                       
\hline                                        
\hline                                        
\noalign{\smallskip}                          
\noalign{\small 
KEY TO REFERENCES. ---
A95     Allen 1995,
C91     Conti 1991,
DM98    Duc \& Mirabel 1998
DRM93   Drissen \etal\ 1993,
ED86    Evans \& Dopita 1986,
GIT98   Guseva \etal\ 1998, 
KS86    Kunth \& Schild 1986,
KSB87   Kunth \etal\ 1987,
MMO91   Masegosa \etal\ 1991,
P99     Pindao \etal\ 1999,
RC93    Robledo-Rella \& Conti 1993,
VC92    Vacca \& Conti 1992,
WO98    Wozniak \etal\ 1998
}
\end{tabular}
}
%\end{flushleft}
\label{ta_cand}
\end{table*}

\end{document}